\documentclass[]{elsarticle}
\usepackage{lineno,hyperref}

\usepackage{graphicx}
\usepackage{colortbl}
\usepackage{float} 
\usepackage{url} 
\usepackage{color}

\usepackage{caption}
\usepackage{subcaption}
\usepackage{setspace}
\usepackage{adjustbox}  

\usepackage{graphicx}
\usepackage{float} 
\usepackage{url} 
\usepackage{color}

\usepackage{graphicx}
\usepackage{amssymb}
\usepackage{latexsym}
\usepackage{amsfonts}
\usepackage{amsmath}
\usepackage{graphics}
\usepackage{setspace}
\usepackage{graphicx}
\usepackage{epstopdf}
\usepackage{color}
\usepackage{float}
\usepackage{wrapfig}
\usepackage{color}
\usepackage{rotating}
\usepackage{array} 
\usepackage{dblfloatfix}
\usepackage{breakurl}

\usepackage{graphics}
\usepackage{setspace}

\usepackage{graphicx}
\usepackage{epstopdf}  
\usepackage{color}
\usepackage{float}
\usepackage{wrapfig}
\usepackage{color}

\usepackage{graphicx}
\usepackage{caption}
\usepackage{subcaption}
\usepackage{setspace}

\usepackage{seqsplit}

\definecolor{blue}{rgb}{0,0,0}
\definecolor{purple}{rgb}{0,0,0}
\definecolor{Mahogany}{rgb}{0,0,0}
\definecolor{olive}{rgb}{0,0,0}

\usepackage{lipsum}
\usepackage{hyperref}
\hypersetup{colorlinks=false}

\usepackage{algorithm}
\usepackage{algorithmic}
\usepackage{morefloats}
\usepackage{textcomp}

\usepackage{caption}
\usepackage{subcaption}
\usepackage{setspace}
\usepackage{adjustbox}  
\usepackage{multirow}    
\definecolor{FULL-SUPPORT}{RGB}{87,157,28}
\definecolor{NO-SUPPORTAT-ALL}{RGB}{204,0,0}
\definecolor{EASY-TO-EXTEND}{RGB}{255,204,0}    
\definecolor{DIFFICULT-TO-EXTEND}{RGB}{255,102,0}
\definecolor{NA}{RGB}{221,221,221}

\definecolor{Computation}{RGB}{255,255,255}
\definecolor{Consent}{RGB}{255,255,255}
\definecolor{Compliance}{RGB}{255,255,255}

\newcommand{\lComp}{\cellcolor{Computation}} 
\newcommand{\lCons}{\cellcolor{Consent}} 
\newcommand{\lCmpl}{\cellcolor{Compliance}}

\definecolor{minimise}{RGB}{255,222,222}
\definecolor{hide}{RGB}{253,228,205}
\definecolor{seperate}{RGB}{250,250,202}
\definecolor{aggregate}{RGB}{219,238,254}
\definecolor{inform}{RGB}{237,215,253}
\definecolor{controlenforce}{RGB}{251,214,253}     
\definecolor{demostrate}{RGB}{228,255,227}

\newcommand{\lMini}{\cellcolor{minimise}} 
\newcommand{\lHide}{\cellcolor{hide}} 
\newcommand{\lSepe}{\cellcolor{seperate}}      
\newcommand{\lAggr}{\cellcolor{aggregate}} 
\newcommand{\lInfo}{\cellcolor{inform}} 
\newcommand{\lCont}{\cellcolor{controlenforce}}      
\newcommand{\lDemo}{\cellcolor{demostrate}}

\usepackage{colortbl}

\journal{Information Sciences}

\usepackage{numcompress}

\begin{document}

\begin{frontmatter}

\title{Designing Privacy-aware Internet of Things Applications}



\author[mymainaddress]{Charith Perera\corref{mycorrespondingauthor}}
\cortext[mycorrespondingauthor]{Corresponding author}
\ead{charith.perera@ieee.org}

\author[mahmoudaddress]{Mahmoud Barhamgi}
\author[OUaddess]{Arosha K. Bandara}
\author[ajmaladdress]{Muhammad Ajmal}
\author[OUaddess]{Blaine Price}
\author[OUaddess]{Bashar Nuseibeh}

\address[mymainaddress]{Cardiff University, United Kingdom}

\address[mahmoudaddress]{Claude Bernard Lyon 1 University, France}
\address[OUaddess]{Open University, United Kingdom}
\address[ajmaladdress]{University of Derby, United Kingdom}

\begin{abstract}
Internet of Things (IoT) applications typically collect and analyse personal data that can be used to derive sensitive information about individuals. However, thus far, privacy concerns have not been explicitly considered in software engineering processes when designing IoT applications. The advent of behaviour driven security mechanisms, failing to address privacy concerns in the design of IoT applications can have security implications.  In this paper, we explore how a Privacy-by-Design (PbD) framework, formulated as a set of guidelines, can help  software engineers integrate data privacy considerations into the design of IoT applications. We studied the utility of this PbD framework by studying how software engineers use it to design IoT applications. We also explore the challenges in using the set of guidelines to influence the IoT applications design process. In addition to highlighting the benefits of having a PbD framework to make privacy features explicit during the design of IoT applications, our studies also surfaced a number of challenges associated with the approach. A key finding of our research is that the PbD framework significantly increases both novice and expert software engineers' ability to design privacy into IoT applications.
\end{abstract}


\end{frontmatter}


\section{Introduction}
\label{sec:Introduction}
The Internet of Things (IoT) \cite{ZMP007}  is a interconnected collection of physical objects or \textit{`things'} that have computing, sensing and actuation capabilities, together with the ability to communicate with each other and other systems to collect and exchange data. The design and development process for IoT applications  is  more  complicated than that for desktop, mobile, or web applications for a number of reasons.  First,  IoT applications  require both software and hardware (e.g., sensors and actuators) to work together across many different types of nodes (e.g., micro- controllers, system-on-chips, mobile phones, miniaturized single-board computers, cloud platforms) with different capabilities under different conditions  \cite{Perera2015a}. Secondly, IoT applications development requires different types of software engineers to work together (e.g., embedded, mobile, web, desktop). The complexity of different software engineering specialists collaborating to combine different types of hardware and software is compounded by the lack of integrated development stacks that support the engineering of end-to-end IoT applications.


Typically, IoT applications  collect and analyse personal data that can be used to derive sensitive information about individuals.  While the misuse of this information can have negative consequences for the individuals concerned, it can also lead to security problems, particularly with advent of new behaviour driven security mechanisms.  For example, implicit authentication techniques \cite{de_luca_touch_2012, shi_implicit_2011} will grant access to systems based on individual behaviour data collected by IoT systems.  This intertwining of security and privacy issues, means that privacy needs to be considered as a key requirement for IoT applications.  However, thus far, privacy concerns have not been explicitly considered (despite isolated solutions \cite{7994599, Xiong}) in software engineering processes when designing and developing IoT applications, partly due to a lack of Privacy-by-Design (PbD) methods for the IoT.   Further, the engineering complexities explained above have forced  software engineers to put most of their efforts towards addressing other challenges such as interoperability and modifiability, resulting in privacy concerns being largely overlooked. \textcolor{blue}{Additionally, a lack of knowledge  about the tangible and intangible benefits of privacy practices have also contributed to privacy challenges being overlooked \cite{Spiekermann2012}.}

We propose to address this issue by providing systematic guidance to help software engineers develop privacy-aware IoT applications.  We build on  earlier work \cite{PrivacyAssessment} which derived a set of privacy  guidelines by examining Hoepman's \cite{Hoepmanraey} eight design strategies and used them to \textit{\textbf{assess}} the privacy capabilities of IoT applications and platforms. This paper integrates these guidelines into a PbD framework that includes a method for applying the guidelines during the IoT application \textit{\textbf{design}} process.  We go on to evaluate how this PbD framework can help software engineers design a number of example IoT applications.

\subsection{Contributions}
\textcolor{blue}{The primary contributions and the scope of this paper are summarised below:}

\begin{itemize}
	\item \textcolor{blue}{We evaluate how a proposed set of privacy guidelines can be used to effectively improve IoT application designs. In support of this, we integrate the guidelines with a method for applying them to propose a PbD framework for IoT applications.}
	\item \textcolor{blue}{Our method is uniquely designed to address the challenges associated with IoT systems, such as their heterogeneity and distributed nature. This is a significant difference from existing PbD frameworks, which focus on more general, high-level principles and design strategies (e.g., \cite{Hoepmanraey}, \cite{Spiekermann2012}).}
	\item \textcolor{blue}{We gain insights into how our framework could help software engineers improve their design of privacy aware IoT applications by identifying and applying privacy protecting features into their designs.}	
	\item \textcolor{blue}{We also explore strengths and weaknesses of our approach as well as challenges in manual application design processes in general. We provide insights on how to address these weaknesses.}
\end{itemize}

\textcolor{blue}{It is important to note that we do not claim our PbD framework is better than any previous work, nor do we claim that applying set of privacy guidelines will eliminate all  privacy risks. To the best of our knowledge, this is one of the first PbD frameworks that explicitly targets IoT application design challenges. Our aim is to maximise software engineers' ability to be aware of and reduce privacy risks at the design phase. We further elaborate the aims of our PbD framework in Section \ref{sec:Guidelines}.}


\subsection{Target Audience}

\textcolor{blue}{We developed our PbD framework as a tool for engineers to help make their designs better in terms through improved privacy awareness.  Therefore, it is important to note that the framework doesn't provide any formal guarantees that IoT systems designed using it will be free of potential privacy problems.  However, we believe software engineers will, at least, be able to apply some privacy guidelines into their design which they would not do otherwise. Mostly, we wanted to help and guide individuals and teams who do not have time, or resources to invest in hiring privacy experts. Completely ignoring privacy issues could cost such small teams a lot it long run as they grow. Later re-factoring is always costly in any software development process. Therefore, our guidelines will help entrepreneurial teams, IoT hackers, hobbyists, etc. to embed privacy protecting features into their IoT application designs at the initial stages without consulting privacy experts. While our guidelines cannot replace privacy experts and consultants in the software engineering process, they can help software engineers to reduce the effort and time needed from privacy experts.}
\\

The paper is structured as follows: Section \ref{sec:Architecting} discusses common IoT architectures and their characteristics. It also briefly introduces the data life cycle phases and their importance when designing privacy into IoT applications. In Section \ref{sec:Motivation}, we present  our motivation through three different use cases. We have used these use cases to evaluate the effectiveness and identify the challenges in designing privacy aware IoT applications. We briefly introduce the PbD framework in Section \ref{sec:Guidelines}. In Section \ref{sec:Evaluations}, we explain the research methodology and evaluate the effectiveness the PbD framework. We discuss our findings and lessons learned in Section \ref{sec:Discussion_and_Lessons_Learned}. Finally, Section \ref{sec:Related_Work} presents the related work and compares our PbD framework with existing approaches. In Section \ref{sec:Conclusion}, we conclude the paper by discussing future directions for our research.

\begin{figure}[!b]
	\centering
	\includegraphics[scale=0.7]{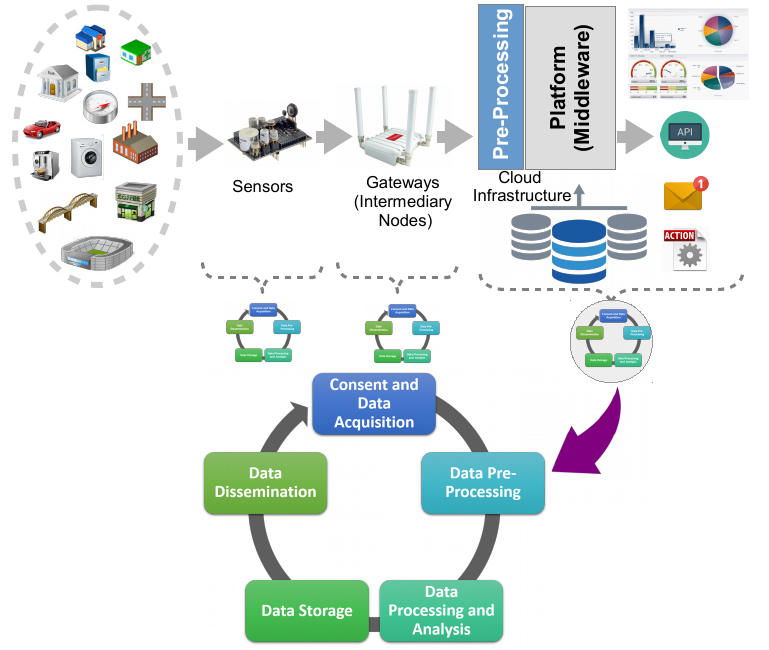}
	\caption{Typical data flow in IoT Applications}
	\label{Figure:DataLifecycle}	
\end{figure}

\section{Internet of Things Software Architecture}
\label{sec:Architecting}

In this section we provide an overview of IoT software architectures from the perspective of how data moves through a typical IoT application.  As illustrated in Figure \ref{Figure:DataLifecycle}, in IoT applications, data moves from sensing devices to gateway devices to the cloud infrastructure  \cite{Perera2015a}. This is the  most common architectural pattern used in IoT application development, which is also called the centralised architecture pattern \cite{roman2013features}. However, there are other patterns such as 1)  collaborative, 2) connected intranet of Things, and 3) distributed IoT \cite{roman2013features}. Even for these other types of architectures, if we consider a single data item, we can observe a data flow analogous to that of the centralised architecture pattern where data moves from edge devices to the cloud through different types of nodes. Therefore, while we use the centralised  IoT architectural pattern to explain our PbD approach in this paper, our approach is agnostic the choice of pattern.


Centralised architectures typically consist of three components: 1) IoT devices, 2) Gateway devices, and 3) IoT cloud platforms (Figure \ref{Figure:DataLifecycle}), each with different computational capabilities. They also have different types of access to energy sources from permanent to solar power to battery power. Each device may also have limitations as to the type of data processing that can be done due to lack of availability of essential knowledge. A typical IoT application would integrate all these different types of devices with different capabilities. It is important to note that different types of privacy protecting measures can be taken on each of these different components based on their  characteristics.

We define a five-phase data life cycle that provides a systematic way of thinking about the data flow in an IoT system for the application of our PbD framework.  Within each device (also called a node), data moves through five data life cycle phases: Consent and Data Acquisition [CDA], Data Preprocessing [DPP], Data Processing and Analysis [DPA], Data Storage [DS] and Data Dissemination [DD]. The CDA phase comprises routing and data collection activities by a given node. DPP describes any type of processing performed on the raw data to prepare it for another processing procedure \cite{pyle1999data}. DPA is, broadly, the collection and manipulation of data items to produce meaningful information \cite{French}. DD is the distribution or transmission of data to an external party. 

All the data life cycle phases are applicable to all nodes in an IoT application, making it possible for software engineers to put in place appropriate mechanisms to protect user privacy. However, based on the decisions taken by engineers, some data life cycle phases in some nodes may not be utilised. For example, a sensor node may utilise the DPP phase to average temperature data. Then, without using either the DPA or DS phases to analyse or store data (due to hardware and energy constraints) the sensor node may push the averaged data to the gateway node using the DD phase.

\section{Example IoT Scenarios}
\label{sec:Motivation}

In this section, we present three use case scenarios, which we also use to evaluate the PbD framework as described in Section \ref{sec:Evaluations}. Each scenario  is presented from a problem owner's perspective, where each problem could be solved by developing an IoT application. More importantly, it should be noted that none of these scenarios explicitly highlight  privacy requirements or challenges. They are primarily focused on explaining functional requirements at a  high level. Later in Section \ref{sec:Guidelines}, we explain how our PbD framework can be used by software engineers to extract additional information from problem owners, that are crucial to design privacy aware IoT applications. 





\subsection{Use case 1: Rehabilitation and Recovery}
\label{sec:Use_case_1}

\textit{\textbf{Summary:}} Robert is a researcher who oversees a number of rehabilitation facilities around the country where patients with physical disabilities are treated and rehabilitated. Robert is interested in collecting and analysing data from sensors worn by patients while they engage in certain activities (e.g., walk using walker, walk using crutches, climbing stairs), in order to guide the patients' recovery processes in a more personalised manner. Robert has an application that is capable of analysing patient data and developing personalised rehabilitation plans. The application monitors the progress and alters the rehabilitation plans accordingly. There is a speciality nurse allocated for each patient in order to monitor the recovery  progress and provide necessary advice when required.

\begin{figure}[h!]
	\centering
	\includegraphics[scale=0.70]{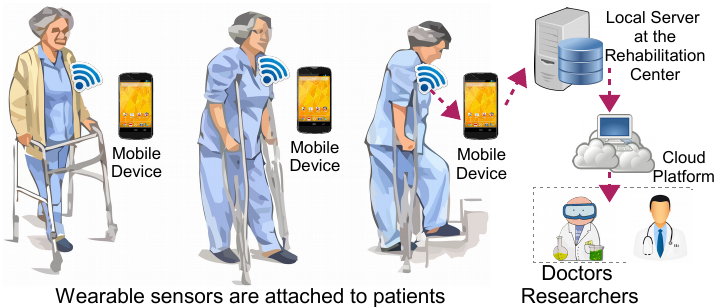}
	\caption{IoT application to support rehabilitation}
	\label{Figure:Usecase2}	
\end{figure}

\subsection{Use case 2: Health and Well-being}
\label{Usecase2}

\begin{figure}[h!]
	\centering
	\includegraphics[scale=0.95]{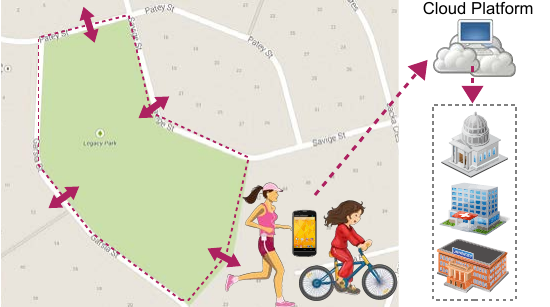}
	\caption{City planning towards health and  well-being}
	\label{Figure:Usecase3}	
\end{figure}

\textit{\textbf{Summary:}} Michael works for the department of public health and well-being. He has been asked to develop a plan to improve the public health in his city by improving the infrastructure that supports exercise and recreational activities (e.g., parks and the paths that supports jogging, cycling, and places for bar exercise, etc.). In order to develop an efficient and effective plan, Michael needs to understand movements of people and several other aspects of their activities. Michael is planning to recruit volunteers in order to gather data using sensors. Michael has an application that is capable of analysing different types of data and recommending possible lifestyle improvements for healthier living. Michael only needs to collect data when the volunteers are within the park premises as illustrated in Figure \ref{Figure:Usecase3}.


\subsection{Use case 3: Amusement Park and Leisure}
\label{sec:Use_case_3}

\textit{\textbf{Summary:}} \textit{TrueLeisure} is a company that operates different types of franchised entertainment attractions. Their amusement parks are located in the United States, United Kingdom, and Australia. These amusement parks are fully owned and operated by franchisees. However, \textit{TrueLeisure} continuously monitors and assesses service quality attributes and several other aspects at each  of the amusement  parks. Jane is a data analyst overseeing the quality assessment tasks  at \textit{TrueLeisure}. She is responsible for continuously monitoring the  service  quality attributes. {Waiting time} is one of the key service quality attributes and  is a key contributory factor to customer satisfaction. Local quality assessment teams continuously measure the crowd {waiting time} of each ride and attraction within their own amusement park. All  visitors use \textit{TrueLeisure}'s theme park mobile app to buy tickets for attractions, further information, tour guide, maps, etc. Jane is interested in the big picture, i.e. she would like to measure the overall waiting time for each ride attraction by combining individual waiting times. Jane will report these measurements to \textit{TrueLeisure} management to guide franchisees on future developments of their theme parks efficiently and effectively.



\section{Privacy-by-Design Framework}
\label{sec:Guidelines}
In each of the example scenarios above, the software engineer would need to perform further analysis to extract explicit privacy requirements that could support the design of privacy enhancing features into the IoT applications that would be developed to deliver the required functionality.  In this section we provide an overview of our PbD framework \cite{PrivacyAssessment} and explain how it could be used to design privacy into IoT applications. \textcolor{blue}{We also explain why guidelines are useful to software engineers and how they relate to other approaches for PbD such as principles, strategies, patterns and tactics.}

\subsection{Why Guidelines (or Heuristics or Check-lists)?}
\label{subsec:Why_Guidelines}
\textcolor{blue}{We use the term \textit{guidelines} as our intention is to guide the software engineers. In general, a guideline  aims to improve or maintain efficiency of a particular process based on  a set of best practices. Guidelines may not be mandatory to follow, but provide recommendations based on experience of dealing with particular problems.  Therefore, the term \textit{heuristics} is also an appropriate term for our guidelines. These techniques rely on using readily accessible, though loosely applicable, information to control problem solving in human beings, machines, and abstract issues \cite{Pearl1984}. Heuristics do not promise to produce perfect or optimal solutions. Finally, the term \textit{check-list} is also appropriate to identify our guidelines.	A check-list is a type of informational aid used to reduce failure by compensating for potential limits of human memory and attention. Our guidelines also aim to reduce human errors by reducing knowledge requirements.}

\textcolor{blue}{Sometimes, guidelines are considered as a less useful approach due to their inherited characteristics such as lack of proof (for consistency or correctness), dependence on the subjective judgement of the follower, lack of rigorous scientific  methods for extracting guidelines, and so on. Despite such weaknesses, guidelines are being used successfully in many domains. The following list showcases some examples where guidelines / heuristics / check-lists are used to address different challenges.}

\begin{itemize}
	\item \textcolor{blue}{Heuristics based usability design and evaluation is widely used in human computer interaction domain \cite{Nielsen1990,Molich1990}.}
	\item \textcolor{blue}{The Information commissioner's office, UK's independent authority set up to uphold information rights in the public interest, use check-lists to guide businesses to prepare themselves for GDPR. \cite{InformationCommissionersOffice}.}
	\item \textcolor{blue}{Surgical Safety Check-list developed for the World Health Organization by Dr. Atul Gawande has been able to reduce mortality by 23\% and all complications by 40\% \cite{Haynes2009}. }
	\textcolor{blue}{Airplane pilots rely upon check-lists to ensure that both routine procedures and emergency responses are handled appropriately \cite{Gawande}.}
\end{itemize}

\textcolor{blue}{The above usages and successes have given us confidence to integrate guidelines into our PbD framework.  The framework combines the guidelines with a method for applying them that avoids the need for individual software engineers to spend time thinking about relevant privacy considerations for their IoT applications themselves. Instead, they can save time and effort by systematically working through the guidelines one by one and checking whether they can apply them. Our node-by-node design methodology also helps manage the complexity of IoT application designs. Guidelines also provide meaningful ways to divide workload among engineers (e.g., each engineer may focus / specialise on addressing a few guidelines) and can be used as a common knowledge base to discuss application designs in teams. Guidelines make the design process comparatively less tiring for engineers as it reduces intensive thinking and knowledge requirements. Guidelines also allow engineers to pause and resume conveniently and keep track of design changes. We acknowledge that guidelines are not perfect and will need to be reviewed and refined over time. However, evidence suggests that guidelines can help to improve effectiveness and efficiency in a range of situations, and in this paper we demonstrate this in the context of privacy aware IoT application design.}

\subsection{Where Guidelines Fit in?}
\begin{figure}[t!]
	\centering
	\includegraphics[scale=0.25]{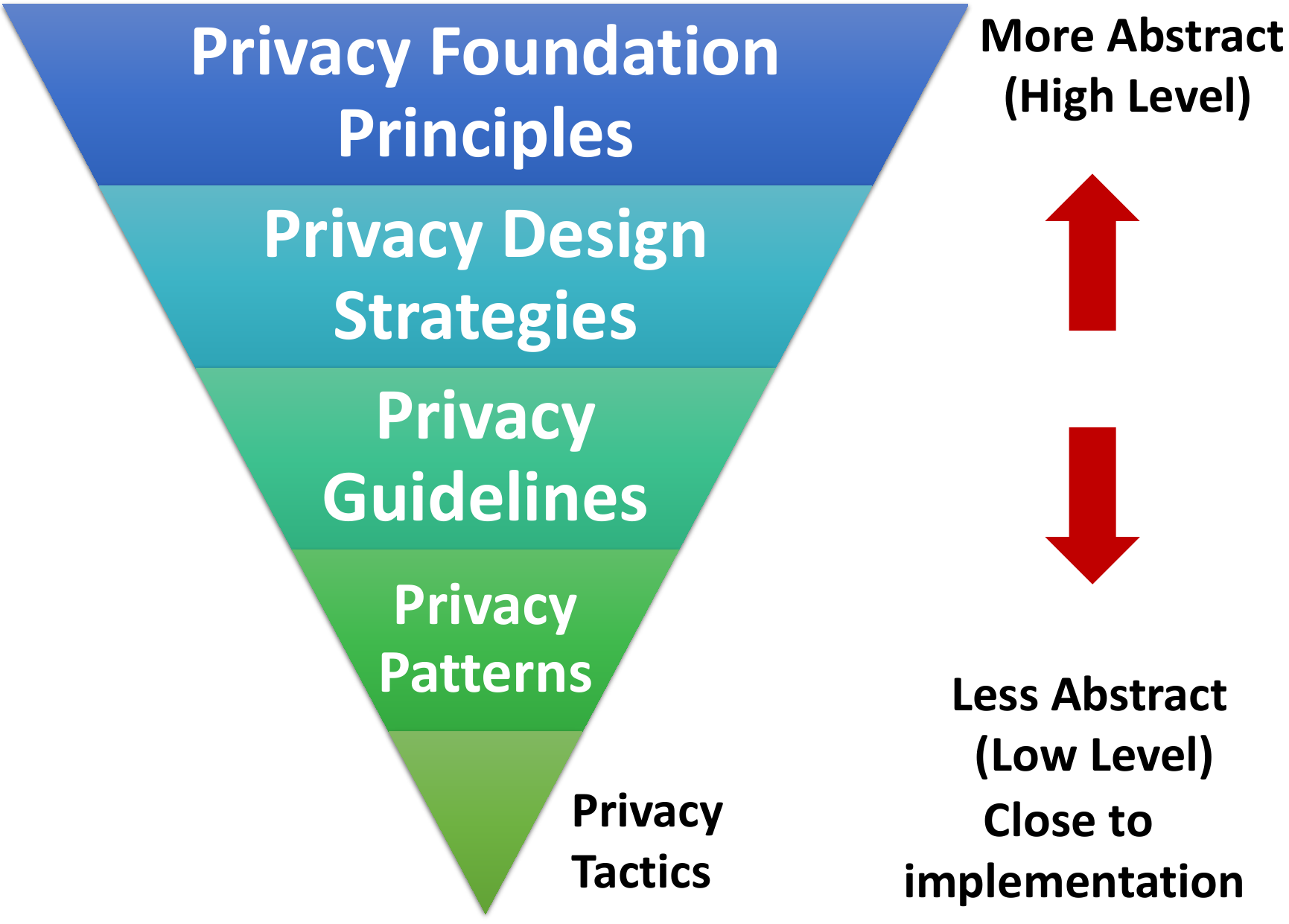}
	\caption{From high level principles to low level tactics:  \textcolor{blue}{Rubinstein and Good \cite{Rubinstein2013}  argue that making a specification or requirement for  software design is to make it concrete, specific, and preferably associated with a metric. The layered approach aims to achieve this in a systematic way.}}
	\label{Figure:BigPicture}	
\end{figure}

\textcolor{blue}{The literature on privacy by design (PbD) techniques uses a number of terms: \textit{principle}, \textit{strategies}, \textit{patterns}, and \textit{tactics}, and in this section we discuss how our concept of PbD guidelines relates to these terms. As shown in Figure \ref{Figure:BigPicture}, principles can be considered to represent high level, more abstract ideas. In contrast, tactics are low level, concrete instructions for implementing solutions in a specific context. Strategies, guidelines and patterns sit in between. This does not mean one type is better or worse than other. Each of these layers have their own strengths and weaknesses. Bottom layer tactics provide specific solutions to specific problems whereas top layer principles provide insights on an overall direction to explore further and solve problems. However, we acknowledge that boundaries between these layers are soft where some principles may be interpreted as strategies and vice-versa.}

\begin{description}
	\item[Principle:] \textcolor{blue}{A principle is a concept or value that is a guide for behaviour or evaluation. Typically, they are very abstract provide an overall direction to follow. Ten Privacy Principles of  Personal Information Protection and Electronic Documents Act (PIPEDA) \cite{MinisterofJustice} and Seven Foundation Privacy by Design principles by Information \& Privacy Commissioner, Canada \cite{Cavoukian} can be identified as examples.}

	\item[Strategies:] \textcolor{blue}{In contrast to principles, strategies are focused on achieving a specific outcome. A design strategy describes a fundamental approach to achieve a certain design goal. Therefore, strategies are more specific in terms of what they aim to achieve. Hoepman's \cite{Hoepmanraey} seven privacy design strategies can be identified as  examples.}
	
	\item[Guidelines:] \textcolor{blue}{The guidelines adopted in this paper break down strategies into a lower-level, concrete set of instructions that a software engineer can follow.}
	
	\item[Patterns:] \textcolor{blue}{Design patterns are useful for making  decisions about the organisation of a software system. A design pattern \textit{``provides a scheme for refining the subsystems or components of a software system, or the relationships between them. It describes a commonly recurring structure of communicating components that solves a general design problem within a particular context.''} \cite{Buschmann1996}. Patterns solve a specific problem but are neutral or have weaknesses with respect to other qualities. In contrast, there are also \textit{`anti-patterns'}. In software engineering, an anti-pattern is a design that may be commonly used but is ineffective or counter productive in practice \cite{Budgen2003}.}
	
	\item[Tactics:] \textcolor{blue}{Patterns are built from tactics (e.g., \textit{if a pattern is a molecule, a tactic is an atom}) \cite{bass_software_2012}. In other terms, patterns package multiple tactics together to solve a specific problem. Tactics help to fine tune patterns and typically they address specific quality attributes and trade-off decisions. Each tactic may have pros and cons. New tactics can be introduced to an existing set in order to address existing weaknesses. However, this could introduce new issues or weaknesses as well. Ideally, we may try different tactics until eventually the side-effects of each tactic become small enough to ignore.}
	
\end{description}

\textcolor{blue}{It is important to note that top three layers (principles, strategies, guidelines) primarily adopt a top-down approach. Typically, we adopt principles, strategies, or guidelines because they suggest good practices and have been historically or logically proven to reduce privacy risks. Typically, they are a blanket solution that aims to eliminate multiple privacy issues at a time (without addressing them individually). In contrast, patterns and tactics focus on solving specific problems. This is more of a bottom-up approach where we try to find solutions to specific privacy problems.}

\textcolor{blue}{Let us explain these layers using an example. This example also highlights the fact that boundaries of these layers can be quite weak at times. \textbf{[Principle]} \textit{``Proactive not Reactive; Preventative not Remedial''} is one of the principles proposed by the Information \& Privacy Commissioner, Canada \cite{Cavoukian}. The official explanation is \textit{``The Privacy by Design (PbD) approach is characterised by proactive rather than reactive measures. It anticipates and prevents privacy invasive events before they happen. PbD does not wait for privacy risks to materialise, nor does it offer remedies for resolving privacy infractions once they have occurred - it aims to prevent them from occurring. In short, Privacy by Design comes before-the-fact, not after.''}.}

\textcolor{blue}{By examining this principle, we may come up with a strategy called \textit{`Minimise'}. \textbf{[Strategy]} Hoepman \cite{Hoepmanraey} describes the \textit{`Minimise'} strategy \textit{``as limiting usage as much as possible by excluding, selecting, stripping, or destroying any storage, collection,retention or operation on personal data, within the constraints of the agreed upon purposes''}. Hoepman's minimise strategy can be identified as a way to follow the \textit{`proactive principle'} (i.e. minimise the amount of data collected as a proactive measure to avoid or reduce potential  privacy violations).} 

\textcolor{blue}{We can further break down the minimise strategy into guidelines. \textbf{[Guideline]} One minimise guidelines is \textit{``Minimise data raw data intake''}. We describe this guideline as \textit{``Whenever possible, IoT applications should reduce the amount of raw data intake. Raw data could lead to secondary usage and privacy violation. Therefore, IoT applications should consider converting (or transforming) raw data into secondary context data.''}\cite{PrivacyAssessment}}

\textcolor{blue}{A privacy pattern can be identified as a low-level design that aims to solve a specific privacy challenge. The relationship between guidelines and patterns may be quite weak as, in most instances, patterns can stand by themselves as problem solving techniques. However, privacy patterns can still be identified as low level designs that help to implement guidelines. Continuing the example, a \textbf{[pattern]} we extract could be \footnote{Detailed discussions about patterns and tactics are outside the scope of this paper.} called \textit{`Online Activity Detector'}. This pattern \textit{extracts orientation (e.g. sitting, standing, walking) by processing accelerometer data and only stores the results (i.e. secondary context) and deletes the raw accelerometer data.}}

\textcolor{blue}{The \textbf{[Tactics]} \textit{`Online Activity Detector'} pattern may compose tactics such as \textit{`average'} and \textit{`periodic delete'} . The \textit{`Average'} tactic may be used to prepare accelerometer data for activity detection. The \textit{`Periodic delete'} tactic may be used to delete data after detection. In some designs \textit{`Periodic delete'} may be replaced with a \textit{`In-memory processing'} tactic which aims to perform the activity detection without writing the data to long-term storage media.}

\begin{table*}[t!]
	\footnotesize 
	\renewcommand{\arraystretch}{1.00}
	\centering
	\caption{Privacy-by-Design Framework}
	\label{Tbl:Analysis}
	\begin{adjustbox}{angle=90}
	\begin{tabular}{|p{6.3cm}lllll|p{0.6cm}|llllllll |l|l| }
		\cline{1-6} \cline{8-15} \cline{17-17}
		
		Guideline                       
		& \begin{sideways}\begin{minipage}[b]{1cm}DA \end{minipage} \end{sideways} 
		& \begin{sideways}\begin{minipage}[b]{1cm}DPP \end{minipage} \end{sideways}  
		& \begin{sideways}\begin{minipage}[b]{1cm} DPA \end{minipage} \end{sideways}  
		& \begin{sideways}\begin{minipage}[b]{1cm} DS \end{minipage} \end{sideways}  
		& \begin{sideways}\begin{minipage}[b]{1cm} DD \end{minipage} \end{sideways} 
		& \begin{sideways}\begin{minipage}[b]{1cm} Capability \end{minipage} \end{sideways}\multirow{4}{*}{}
		& \begin{sideways}\begin{minipage}[b]{1.6cm}Minimise \end{minipage} \end{sideways} 
		& \begin{sideways}\begin{minipage}[b]{1.6cm}Hide \end{minipage} \end{sideways}  
		& \begin{sideways}\begin{minipage}[b]{1.6cm} Separate \end{minipage} \end{sideways}  
		& \begin{sideways}\begin{minipage}[b]{1.6cm} Aggregate \end{minipage} \end{sideways}  
		& \begin{sideways}\begin{minipage}[b]{1.6cm} Inform \end{minipage} \end{sideways}   
		& \begin{sideways}\begin{minipage}[b]{1.6cm} Control \end{minipage} \end{sideways}  
		& \begin{sideways}\begin{minipage}[b]{1.6cm} Enforce \end{minipage} \end{sideways}  
		& \begin{sideways}\begin{minipage}[b]{1.6cm} Demonstrate \end{minipage} \end{sideways} 
		&& \begin{sideways}\begin{minipage}[b]{2.0cm} Privacy Risks \end{minipage} \end{sideways} \\ \cline{1-6} \cline{8-15} \cline{17-17}
		\arrayrulecolor{NA}
		\lMini 1-Minimise data acquisition                &  \checkmark      &    \checkmark       &                   &              &              &\lComp& \checkmark  &  \checkmark&            &            &            &            &            &       & & $\otimes$ $\circleddash$    \\   \cline{1-6} \cline{8-15}  \cline{17-17}
		\lMini 2-Minimise number of data sources          &  \checkmark      &                     &                   &              &              &\lComp& \checkmark  &            &            &            &            &            &            &       &&  $\otimes$ $\circleddash$\\  \cline{1-6} \cline{8-15} \cline{17-17}
		\lMini 3-Minimise raw data intake                 &  \checkmark      &    \checkmark       &                   &              &              &\lComp& \checkmark  &            &            &  \checkmark&            &            &            &     &&  $\otimes$ $\circleddash$  \\  \cline{1-6} \cline{8-15} \cline{17-17}
		\lMini 4-Minimise knowledge discovery             &                  &                     &   \checkmark      &              &              &\lComp& \checkmark  &            &            &            &            &            &            &      && $\otimes$  \\  \cline{1-6} \cline{8-15} \cline{17-17}
		\lMini 5-Minimise data storage                    &                  &                     &                   &   \checkmark &              &\lComp& \checkmark  &            &            &            &            &            &            &      && $\otimes$  $\circleddash$\\  \cline{1-6} \cline{8-15} \cline{17-17}
		\lMini 6-Minimise data retention period           &                  &                     &                   &   \checkmark &              &\lComp& \checkmark  &  \checkmark&            &            &            &            &            &      && $\otimes$ $\circleddash$ \\  \cline{1-6} \cline{8-15} \cline{17-17}
		\lHide 7-Hidden data routing                      &  \checkmark      &                     &                   &              &  \checkmark  &\lComp&             &  \checkmark&            &            &            &            &            &      &&  \textcolor{white}{$\otimes$} $\circleddash$ \\  \cline{1-6} \cline{8-15} \cline{17-17}
		\lHide 8-Data anonymisation                       &  \checkmark      &    \checkmark       &   \checkmark      &              &  \checkmark  &\lComp&             & \checkmark &            &            &            &            &            &      && $\otimes$  $\circleddash$\\  \cline{1-6} \cline{8-15} \cline{17-17}
		\lHide 9-Encrypted data communication             &  \checkmark      &                     &                   &              &  \checkmark  &\lComp&             & \checkmark &            &            &            &            &            &      && \textcolor{white}{$\otimes$} $\circleddash$  \\  \cline{1-6} \cline{8-15} \cline{17-17}
		\lHide 10-Encrypted data processing               &                  &    \checkmark       &   \checkmark      &              &              &\lComp&             &\checkmark  &            &            &            &            &            &      && \textcolor{white}{$\otimes$} $\circleddash$  \\  \cline{1-6} \cline{8-15} \cline{17-17}
		\lHide 11-Encrypted data storage                  &                  &                     &                   &              &              &\lComp&             & \checkmark &            &            &            &            &            &      && \textcolor{white}{$\otimes$} $\circleddash$  \\  \cline{1-6} \cline{8-15} \cline{17-17}
		\lHide 12-Reduce data granularity                 &  \checkmark      &    \checkmark       &   \checkmark      &              &   \checkmark &\lComp& \checkmark  & \checkmark &            &            &            &            &            &      &&  $\otimes$ \\  \cline{1-6} \cline{8-15} \cline{17-17}
		\lHide 13-Query answering                         &                  &                     &                   &              &   \checkmark &\lComp& \checkmark  & \checkmark &            &            &            &            &            &      &&  $\otimes$ \\  \cline{1-6} \cline{8-15} \cline{17-17}
		\lHide 14-Repeated query blocking                 &                  &                     &                   &              &   \checkmark &\lComp&  \checkmark & \checkmark &            &            &            &            &            &     &&  $\otimes$  \\  \cline{1-6} \cline{8-15} \cline{17-17}
		\lSepe 15-Distributed data processing             &                  &                     &   \checkmark      &              &              &\lComp&             &            & \checkmark &            &            &            &            &     &&  \textcolor{black}{$\otimes$} $\circleddash$  \\  \cline{1-6} \cline{8-15} \cline{17-17}
		\lSepe 16-Distributed data storage                &                  &                     &                   &  \checkmark  &              &\lComp&             &            & \checkmark &            &            &            &            &     &&  $\otimes$ $\circleddash$  \\  \cline{1-6} \cline{8-15} \cline{17-17}
		\lAggr 17-Knowledge discovery based aggregation   &  \checkmark      &    \checkmark       &   \checkmark      &  \checkmark  &  \checkmark  &\lComp&             &            &            &  \checkmark&            &            &            &     && $\otimes$   \\  \cline{1-6} \cline{8-15} \cline{17-17}
		\lAggr 18-Geography based aggregation             &  \checkmark      &    \checkmark       &   \checkmark      &  \checkmark  &   \checkmark &\lComp&             &            &            &  \checkmark&            &            &            &     &&  $\otimes$  \\  \cline{1-6} \cline{8-15} \cline{17-17}
		\lAggr 19-Chain aggregation                       &  \checkmark      &    \checkmark       &   \checkmark      &  \checkmark  &   \checkmark &\lComp&             &            &            &  \checkmark&            &            &            &     &&  $\otimes$  \\  \cline{1-6} \cline{8-15} \cline{17-17}
		\lAggr 20-Time-Period based aggregation           &  \checkmark      &    \checkmark       &   \checkmark      &  \checkmark  &   \checkmark &\lComp&             &            &            &  \checkmark&            &            &            &      &&  $\otimes$ \\  \cline{1-6} \cline{8-15} \cline{17-17}
		\lAggr 21-Category based aggregation              &  \checkmark      &    \checkmark       &   \checkmark      &  \checkmark  &   \checkmark &\lComp&             &            &            &  \checkmark&            &            &            &      && $\otimes$  \\  \cline{1-6} \cline{8-15} \cline{17-17}
		\lInfo 22-Information Disclosure                  &  \checkmark      &    \checkmark       &   \checkmark      &  \checkmark  &   \checkmark &\lCons&             &            &            &            & \checkmark &            &            &\checkmark   &&  $\otimes$\\  \cline{1-6} \cline{8-15} \cline{17-17}
		\lCont 23-Control                                 &  \checkmark      &    \checkmark       &   \checkmark      &  \checkmark  &   \checkmark &\lCons&             &            &            &            &            & \checkmark & \checkmark &       && $\otimes$ \\  \cline{1-6} \cline{8-15} \cline{17-17}
		\lDemo 24-Logging                                 &  \checkmark      &    \checkmark       &   \checkmark      &  \checkmark  &   \checkmark &\lCmpl&             &            &            &            &            &            &  		   &\checkmark     && $\otimes$ $\circleddash$   \\  \cline{1-6} \cline{8-15} \cline{17-17}
		\lDemo 25-Auditing                                &  \cellcolor{NA}  &   \cellcolor{NA}    &   \cellcolor{NA}  &\cellcolor{NA}&\cellcolor{NA}&\lCmpl&             &            &            &            &            &            &  		   &\checkmark     &&    \\  \cline{1-6} \cline{8-15} \cline{17-17}
		\lDemo 26-Open Source                             &  \cellcolor{NA}  &   \cellcolor{NA}    &    \cellcolor{NA}  &\cellcolor{NA}&\cellcolor{NA}&\lCmpl&             &            &            &            &            &            &  		   &\checkmark      &&   \\  \cline{1-6} \cline{8-15} \cline{17-17}
		\lDemo 27-Data Flow Diagrams                      &  \cellcolor{NA}  &   \cellcolor{NA}    &   \cellcolor{NA}  &\cellcolor{NA}&\cellcolor{NA}&\lCmpl&             &            &            &            &            &            &  		   &\checkmark     &&    \\  \cline{1-6} \cline{8-15} \cline{17-17}
		\lDemo 28-Certification                           &  \cellcolor{NA}  &   \cellcolor{NA}    &   \cellcolor{NA}  &\cellcolor{NA}&\cellcolor{NA}&\lCmpl&             &            &            &            &            &            &  		   &\checkmark      &&   \\  \cline{1-6} \cline{8-15} \cline{17-17}
		\lDemo 29-Standardisation                         &  \cellcolor{NA}  &   \cellcolor{NA}    &   \cellcolor{NA}  &\cellcolor{NA}&\cellcolor{NA}&\lCmpl&             &            &            &            &            &            &  		   &\checkmark     &&    \\  \cline{1-6} \cline{8-15} \cline{17-17}
		\lDemo 30-Compliance                              &  \cellcolor{NA}  &   \cellcolor{NA}    &   \cellcolor{NA}  &\cellcolor{NA}&\cellcolor{NA}&\lCmpl&             &            &            &            &            &            &  		   &\checkmark    &&  $\otimes$ $\circleddash$  \\  \cline{1-6} \cline{8-15}  \cline{17-17}
		
		\multicolumn{17}{l}{ Risk Types: Secondary Usage ($\otimes$), Unauthorised Access ($\circleddash$)}\\  
	\end{tabular}
\end{adjustbox}
\end{table*}

\subsection{Overview of the Guidelines}
For ease of reference, we present an overview of our privacy guidelines in Table \ref{Tbl:Analysis}.  These guidelines are based on Hoepman's \cite{Hoepmanraey} privacy strategies, which we determined to be the most appropriate starting point for developing a more detailed set of PbD guidelines for IoT applications. \textcolor{blue}{The guidelines were compiled by using the structured-case research method \cite{Carroll2000}, which is often used for building theory in information systems research.} A more detailed explanation on each of the guidelines and reasoning behind the extraction of each guideline is presented in \cite{PrivacyAssessment}. 

The guidelines allow software engineers to customise them as needed to suit their IoT applications. For example, certain applications will require aggregation of data from different sources to discover a certain new knowledge (i.e. new pieces of information). Such approaches are not discouraged as long as data is acquired through proper consent acquisition processes. However, IoT applications, at all times, should take all possible measures to achieve their goals with a minimum amount of data. This means that out of the eight privacy design strategies proposed by Hoepman \cite{Hoepmanraey}, minimisation is the most important strategy. 

In our previous work \cite{PrivacyAssessment}, we identified two major privacy risks, namely, secondary usage ($\otimes$) and unauthorised access ($\circleddash$) that would arise as consequences of not following the guidelines. Secondary usage refers to the use of collected data for purposes that were not initially consented to by the data owners \cite{Lowrance2003}, which can lead to privacy violations. Unauthorised access is when someone breaches the confidentiality of the data during any phase of the data life cycle by gaining without proper authorisation. The above symbols above are used to denote which threat is relevant to each guideline. In Table \ref{Tbl:Analysis}, privacy guidelines are colour coded based on the primary privacy design strategy that they belong to. However, it is important to note that some guidelines may belong to multiple design strategies. For example, \textbf{(Guidelines 6)} \textit{minimise data retention period} can primarily be identified as a minimise strategy, but it can also be classified as a hide strategy as it reduces the period for which data is visible.

\subsection{Use of Privacy-by-Design Framework}
\label{sec:example}
The objective of the proposed PbD framework is to help software engineers to ask the right questions regarding privacy protection when designing IoT applications and their architectures. Our approach integrates privacy guidelines into a framework that includes a method for engineers to start thinking about privacy and incorporate privacy features into IoT application designs. A piece of software is designed to solve a problem. Sometimes, a problem may be identified by a person who is affected by the problem (e.g., Robert, Michael or Jane in our motivating scenarios). At other times, a third party company may identify a generic problem that affects many other people (e.g., Enterprise resource planning solutions). This type of software engineering is common in the IoT domain as well.  Some IoT solutions are generic middleware platforms that can be used to build end to end applications. Others are complete IoT applications that aim to solve a specific problem \cite{Perera2015a,PereraIEEEAccess}.

However, problem owners mainly focus  on the requirements that would help to solve their problem \cite{bass_software_2012}, ignoring privacy considerations. Therefore,  privacy requirements are largely overlooked when designing software architectures for IoT applications.  The PbD framework allows both problem owners and software engineers to sit together and discuss the problem and incorporate privacy protecting measures into IoT application designs. 

In section \ref{sec:Motivation}, we presented three use case scenarios. For each scenario, we have a problem owner's expectation and a brief set of requirements. There is no explicit reference to privacy protecting measures. We assume, additional  information can only be gathered through questioning the problem owners and domain experts. In the studies reported later in this paper, we simulated such discussions between the problem owners (i.e, represented by ourselves, the researchers) and the software engineers (i.e., represented by the study participants).  Our hypothesis was that the PbD framework will help software engineers to ask questions from both problem owners and domain experts in order to extract detailed requirements that could be used to design privacy into IoT applications.

To illustrate how this might work in practice, let us revisit the scenario presented in section \ref{sec:Use_case_1} and use our  PbD framework to extract privacy requirements for designing a privacy-aware IoT application. 

\textbf{Guideline 1} leads software engineers to ask the question: what types and quantities of data are required to achieve the Robert's objective?  In our scenario the problem owner responds as follows:

\textit{Robert collects data using wearable sensor kits. The collected data types are pulse, oxygen in blood (SPO2), airflow (breathing), body temperature, electrocardiogram (ECG), glucometer, galvanic skin response (GSR-sweating), blood pressure (sphygmomanometer), patient activity (accelerometer) and muscle / eletromyography sensor (EMG). Accelerometer is used to derive patient activity.  In addition to the sensor data, weather information such as temperature, humidity are also important for the Robert's research. Patients' mobile phones GPS sensors and weather APIs are used to collect such information. The data collection sampling rate is expected to be 30 seconds. Data is only required to be collected when patients are performing either one of the monitored activities (i.e. walking with walker or crutches, or climbing stairs). } 

Based on this information the software engineer can decide not to acquire any other types of data and also design appropriate sampling rate controls into the application.  This will have the effect of minimising data acquisition and reducing the risk of both secondary usage and unauthorised access to private data.

In a similar fashion, guidelines 3, 5, 20 and 21 would lead a software engineer to ask questions such as: what type of data is required in raw format and what type of information can be aggregated in order to reduce privacy risks?. As a result, the following information may be gathered.

\textit{Robert requires oxygen in blood (SPO2), airflow (breathing), body temperature data types in raw format which need to be accurate. The data collection sampling rate is expected to be five seconds. In contrast, other data items can be aggregated into averaged values (e.g., aggregated over two minutes). }


Similar guidelines based questioning can be used to extract privacy requirements which the software engineers can use to systematically design privacy into the IoT application. Due to space limitations, we don't detail all the questions that could be asked in relation to the scenario.  Instead, below we provide the information that could be acquired using our PbD approach by annotating a detailed description of the scenario with references to the relevant PbD guidelines at the end of each statement.

\textit{The sensor kit is expected to push data to the patient's mobile phone using Bluetooth. The mobile phone pushes data to the rehabilitation centre's local server using Wi-Fi. The local server pushes data to the cloud IoT platform. Patients come to the rehabilitation centre 3 days a week in order to perform the tasks assigned to them. Another 3 days they perform the task at their homes. The smart phone is expected to push data to the local server at the end of each day (\textbf{\textit{Guideline 6}}). However, if the patients perform their tasks at home, data need to be kept stored on the mobile until the next time they visit the rehabilitation centre (\textbf{\textit{Guideline 6}}). The speciality nurses monitor the progress and advise the patients on weekly basis. The speciality nurses' responsibility is to make sure that the patient are performing the tasks as assigned by the recommendation system and assists patients if they have any difficulties in following the assigned tasks and schedules. Robert is required to analyse data every six months in order to understand the how to improve the rehabilitation processes in a personalized manner (\textbf{\textit{Guideline 6}}). For long term data analysis purposes, Robert's application stores data after averaging over five minutes (\textbf{\textit{Guideline 6}}).}

\textit{Robert's application requires averages over five minutes when patients are performing their tasks (\textbf{\textit{Guideline 20}}). Patient data can be anonymized (\textbf{\textit{Guideline 8}}). Data storage in both mobile device, local server and Robert's cloud server should store data in encrypted form (\textbf{\textit{Guideline 11}}). End-to-end encryption can be used to secure the data communication (\textbf{\textit{Guideline 9}}). Robert does not require the exact locations where patients may have performed the activities. The requirement is to acquire the weather parameters such as temperature, humidity, etc. Therefore, location data can be abstracted without affecting the accuracy of the data (\textbf{\textit{Guideline 12}}). In this IoT application, data processing and storage happens in three different nodes, namely, 1) patient phone, 2) local server, and 3) Robert's cloud server (\textbf{\textit{Guideline 15 and 16}}).}

The above example illustrates how the PbD guidelines could be used to extract additional information regarding a use case which enables software engineers to design appropriate privacy enhancing features into their IoT applications.  In order to evaluate the effectiveness of our PbD framework, we developed similar detailed requirement descriptions for each of the use case scenarios, which we have omitted here due to space limitations.  It is important to note that not all privacy guidelines are relevant to all IoT applications. In Table \ref{Tbl:Privacy_Requirement_Extraction}, we summarise which privacy guidelines are relevant to each scenario. 

\begin{table}
	\caption{Relevant Privacy Requirements for Each Use Case Scenario}
	\small
	\begin{center}
		\begin{tabular}{l c c c}
			
			Guideline ($\downarrow$) \  Use Case Number ($\rightarrow $)&  1 &  2 & 3 \\ \hline
			1-Minimise data acquisition                 & \checkmark & \checkmark & \checkmark \\ 
			2-Minimise number of data sources           & --         & \checkmark & -- \\ 
			3-Minimise raw data intake                  & \checkmark & \checkmark & \checkmark \\ 
			5-Minimise data storage                     & \checkmark & \checkmark & \checkmark \\ 
			6-Minimise data retention period            & \checkmark & \checkmark & \checkmark \\ 
			7-Hidden data routing                       & \checkmark & \checkmark & \checkmark \\ 
			8-Data anonymisation                        & \checkmark & \checkmark & \checkmark \\ 
			9-Encrypted data communication              & \checkmark & \checkmark & \checkmark \\ 
			11-Encrypted data storage                   & \checkmark & \checkmark & \checkmark \\ 
			12-Reduce data granularity                  & \checkmark & \checkmark & \checkmark \\ 
			15-Distributed data processing              & \checkmark & \checkmark & \checkmark \\ 
			16-Distributed data storage                 & \checkmark & \checkmark & \checkmark \\ 
			18-Geography based aggregation              & --         & --         & \checkmark \\ 
			20-Time-Period based aggregation            & \checkmark & \checkmark & \checkmark \\ 
			21-Category based aggregation               & \checkmark & \checkmark & \checkmark \vspace{5pt}\\

			& 13 & 14 & 14 \\ 
		\end{tabular}
	\end{center}
	\label{Tbl:Privacy_Requirement_Extraction}
	\vspace{-12pt}
\end{table}

\section{Evaluation}
\label{sec:Evaluations}

This section explains how we evaluated our PbD framework, together with our research methodology. Specifically, our evaluation is based on the following two studies:
\begin{enumerate}
	\item \textcolor{olive}{\textbf{Study 1 (Primary)}: [Interview based] This was our primary study in which we tested our main hypothesis: \textit{`Can the proposed PbD framework guide software engineers with varied levels of experience to design IoT applications that are more privacy-aware than they would do otherwise?'} Additionally, we explored engineers' perception of each guideline, their usefulness, and applicability in different IoT use case scenarios - collectively referring to this as the engineers' \textit{privacy mindset}. The study was administered by a researcher and focused on both quantitative (for hypothesis testing) and qualitative data. }
	
	\item \textcolor{olive}{\textbf{Study 2 (Secondary)}: [Online activity based] This was a self administered online study. In this study, we explored the engineers' privacy mindset with respect to each guideline. In contrast to Study 1, here we used an anonymous, informal, and relaxed  methodology using a self administered online activity that could be completed over a 3-day period. We used this study to strengthen our findings from Study 1 as well as to reach theoretical saturation\footnote{Theoretical saturation is the phase of qualitative data analysis in which the researcher has continued sampling and analysing data until no (or very minimal) new data appear \cite{Lewis-Beck2004}}. In this study, we mainly focused on qualitative data (though we present some quantitative aspects).}
\end{enumerate}

For each study, we first explain the aims of the study followed by a description of the participant recruitment strategy and sample size.  Finally we describe the procedures followed at each step of the study. In adopting this approach, we were partially inspired by the evaluation strategies used by comparable techniques, particularly the evaluation methodology used for LINDDUN \cite{LINDDUN}, including adopting a use case  based evaluation technique.

\subsection{Study 1 (Primary) - Interview-based}
\label{sec:studey1}	

\subsubsection{{Purpose}}
The purpose of this  study is to explore how our PbD framework can help  software engineers to design privacy-aware IoT applications. Through user studies, using quantitative and qualitative data analysis, we aimed to answer following three  questions that explore the effectiveness  of the proposed PbD framework. We discuss these questions later in this section.

%
%

\begin{itemize}
	\item {Can the proposed PbD framework guide less experienced (novice) software engineers to design IoT applications that are more privacy-aware than they would do otherwise?}
	
	
	\item {Can the proposed PbD framework guide more experienced (expert) software engineers to design IoT applications that are more privacy-aware than they would do otherwise?}
	
	\item {Out of novice and expert software engineers, who would benefit most from the proposed PbD framework? or in other words, does the level of software engineering expertise matter when it comes to incorporating privacy protection features into IoT application designs?}
\end{itemize}

In the first two questions above, we consider the design of an IoT application to be more privacy-aware if the designer considers a greater number of privacy concerns to incorporate appropriate privacy protecting features.  We measure this in terms of the number of privacy guidelines considered by the study participants when designing the example IoT applications.

\subsubsection{Recruitment and Remuneration}

In total, we recruited 10 participants for the study of which five were novice software engineers and five were expert software engineers. A participant was classified as a novice if they had less than three years of experience (full-time) in a software engineering role (design or development). Participants with more than three years of experience (design or development), were considered to be experts.  We adopted an opportunistic sampling technique and participants were recruited from the staff and student populations across two universities in the United Kingdom.  No criteria other than software engineering experience was considered when recruiting participants. We collected demographic information such as age, highest education qualification, and the number of years in a software engineering role. Each participant was compensated with shopping vouchers valued at GBP 20. There were no failure criteria as long as the participant attend the data collection session of the study.  The study design was reviewed and approved by our institution's Human Research Ethics Committee. \textcolor{blue}{Table \ref{Tbl:ParticipantsDemographics} summarises  the demographic information about the participants. We have labelled them E1-E5 (Expert) and N1-N5 (Novice) and consider them to be independent cases for the purposes of our qualitative analysis process.}

\begin{table*}[b!]
	\centering
	\footnotesize
	\caption{Demographics of Study 1 (Primary Study)}
	\label{Tbl:ParticipantsDemographics}
	\begin{tabular}{llccc}
		\multicolumn{1}{l}{ID}   & 
		\multicolumn{1}{c}{Age} & 
		\begin{sideways}\begin{minipage}[b]{2cm}{Highest \\Qualification (ICT)} \end{minipage} \end{sideways}   & 
			\begin{sideways}\begin{minipage}[b]{2cm}{Years of \\Experience} \end{minipage} \end{sideways}   & 
			\begin{sideways}\begin{minipage}[b]{2cm}{Area of \\Experience} \end{minipage} \end{sideways} 
		\\ \hline 
		E1 (Male)                &  20-29                   & MSc          &   4 (Expert)         &  Desktop, Mobile, Web                                     \\
		E2 (Female)              &  30-39                   & PG(Diploma)  &   8 (Expert)         &  Mobile, web, system integration                          \\
		E3 (Female)              &  30-39                   & MSc          &   8 (Expert)         &  Embedded, Textile Design, wearable   \\
		E4 (Male)                &  40-49                   & BSc          &   10 (Expert)        &  Data Science                                             \\
		E5 (Male)                &  20-29                   & BSc          &   6.5 (Expert)       &  Desktop, Mobile, Web                                     \\
		N1 (Male)                &  30-39                   & PhD          &   3 (Novice)         &  Signal Processing                                        \\
		N2 (Male)                &  30-39                   & MSc          &   2.5 (Novice)       & Desktop                                                            \\
		N3 (Male)                &  20-29                   & BSc          &   3 (Novice)         &  Desktop                                                  \\
		N4 (Male)                &  30-39                   & MSc          &   1 (Novice)         &  Desktop                                                \\	
		N5 (Male)                &  30-39                   & MSc          &   3 (Novice)         &  Web                                      
	\end{tabular}
\end{table*}

\subsubsection{Procedure}

\begin{figure}[!b]
	\centering
	\vspace{-0.33cm}
	\includegraphics[scale=0.45]{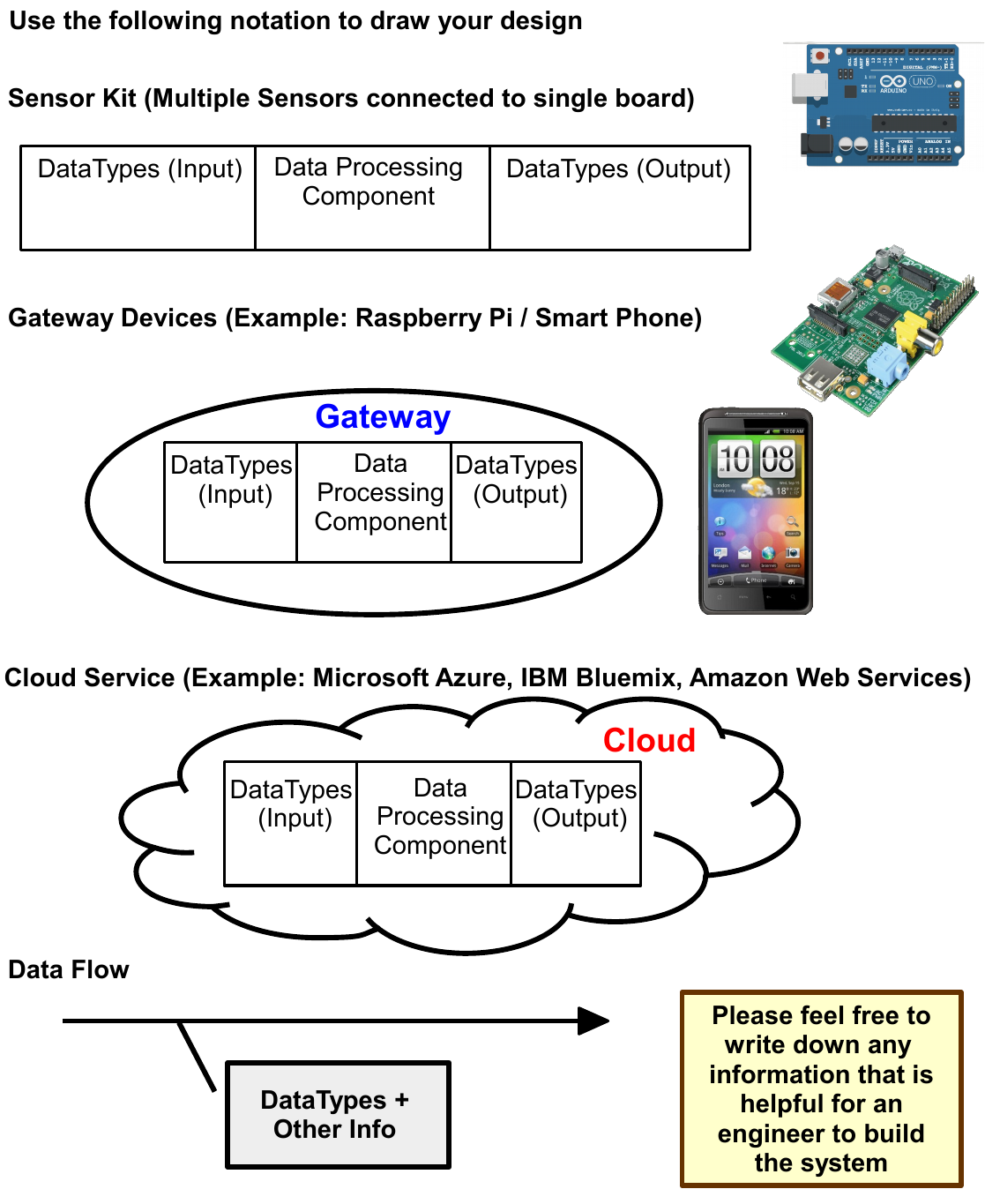}
	\caption{Notations to be used in IoT application Design}
	\label{Figure:Notation}	
\end{figure}

All the data collection sessions were carried out as 1-to-1 lab-based observational studies \cite{Sharp2015}. The principal investigator (PI) acted as the facilitator as well as the observer during each of the sessions. The duration of each session was 1.5 hours. At the beginning of the each session, participants were given the consent form to sign off and brief demographic information was collected. We audio recorded all the discussions between the participants and the PI for qualitative analysis purposes. Next, participants were given an instruction sheet, as shown in Figure \ref{Figure:Notation}, that comprised a set of example notations that could be used to illustrate the design of the IoT applications.  Participants were reassured that adherence to the notation was not essential.

We divided the rest of the study into three rounds, first with no guidance to consider privacy or reference to the PbD framework (Round 1), then with a prompt to consider privacy requirements for the use cases but no reference to the PbD framework (Round 2), and finally using the PbD framework (Round 3). However, this segmentation was only used to structure the discussions and observations and none of the rounds were formally acknowledged or identified during the sessions.

\textbf{Round 1 (NoPrivacy) -} \textit{IoT application design without any guidance to consider privacy or reference to the PbD guidelines}:  It is important to note that we informed the participants that this is an IoT application design study, without making any reference to privacy. This was done with the expectation that participants would be unbiased and follow their natural process for designing an IoT application. We gave them separate A4 sheets to draw their IoT application designs with respect to each use case. They were briefed about the notations they could use, but we did not restrict them to any particular notation as long as their designs were understandable and clearly annotated.

The participants were asked to design IoT applications to satisfy the requirements of each the scenarios presented in Section \ref{sec:Motivation}.  Initially the participants worked from the summary descriptions provided in this paper but the PI was prepared to provide more detailed information, similar to that presented in Section \ref{sec:example} if the participant explicitly asked any related questions.  We designed the study to simulate a conversation between a software engineer and a problem owner, where the engineer is trying to elaborate the requirements and design the architecture of the IoT application. 

\begin{sidewaysfigure}
	\centering
	\includegraphics[width=\columnwidth]{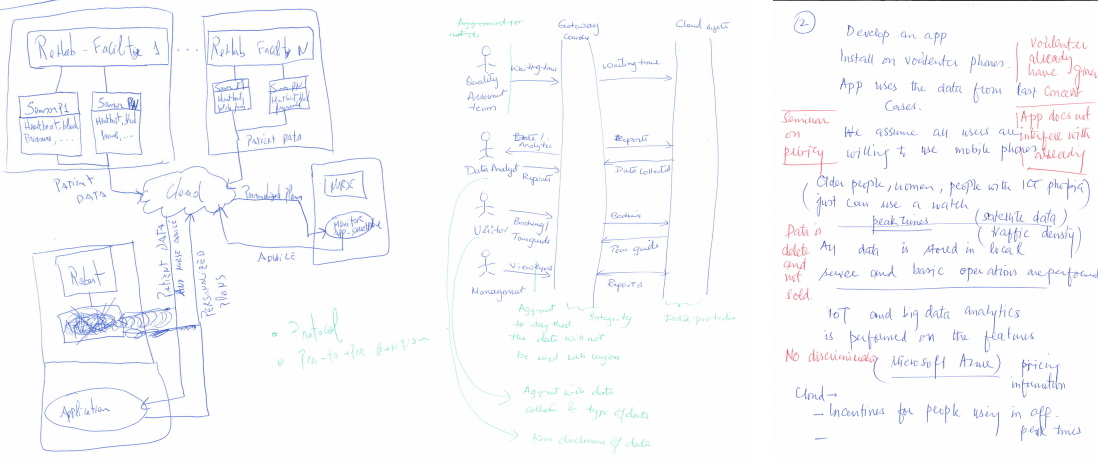}
	\caption{Sample IoT application designs that illustrate a variety of approaches used by participants to express their high-level designs.  In addition to block diagram notations based on the examples we provided, participants used sequence diagrams, pictorial diagrams and detailed text descriptions as illustrated above. }
	\label{Figure:sample_designs}	
\end{sidewaysfigure}

We encouraged participants to ask as many question as possible about the case studies and application requirements. This means that participants could have asked any question regarding privacy requirements if they wanted to. Some of the commonly asked questions are discussed later in this paper. We gave them 50 minutes to complete the IoT application designs for the three use cases provided. However, the time limit was only a guide to the participants and was not enforced. The actual time taken for each study varied based on the time taken by the participants on each phase. So the actual total time varied between 1 hour and 15 minutes to 2 hours. We always allowed each participant to naturally progress through their designs without rushing them through each phase.  After the designs were completed, we asked the participants to explain their designs and briefly justify their design decisions.

\textbf{Round 2 (WithPrivacy)-} \textit{IoT application design with guidance to consider privacy but without privacy guidelines}:Next, we gave participants a ten minute introduction on privacy. In order to achieve consistency, accuracy, and a well recognised description of privacy and related challenges, we selected two videos\footnote{What Is Privacy? (youtube.com/watch?v=zsboDBMq6vo)} \footnote{Big Data (youtube.com/watch?v=HOoKhnvoYkU)}  from YouTube produced and published by \textit{Privacy International} (www.privacyinternational.org).  The objective of showing these videos to each participant was to provoke them to think about privacy and help them to recall their past experiences and knowledge of dealing with privacy issues. This was intended to help them with the next task. It is important to note that we did not provide any additional material on privacy at this stage. 

Next, we asked the participants to refine their previous IoT application designs further to protect user privacy. Similar to the previous round, questions were welcomed.  We gave the participants 20 minutes to refine the IoT application designs for the three use cases provided.  For Round 2, they wrote in a different colour to round 1,  which enabled us to distinguish the design activities from each round clearly. After the revisions were made, we asked the participants to explain their revised designs and how they improved privacy protection.

\textbf{Round 3 (WithPbDGuidelines) -} \textit{IoT applications design with privacy guidelines}: Finally, we gave participants an introduction on the PbD guidelines and how to use them. We asked the participants to refine their previous IoT application designs to protect user privacy. Similar to the previous round, questions were welcomed.  We gave the participants 20 minutes to enhance the privacy features of their IoT application designs for the three use cases provided. After the revisions were made, we asked the participants to explain their revised designs and how they improve privacy protection. Once completed, we collected the IoT application designs produced by the participant. Some sample application designs produced by participants are presented in Figure \ref{Figure:sample_designs}.

\subsection{Study 2 (Secondary) - Online Activity-based}

\subsubsection{Purpose}
\textcolor{olive}{Study 1 was conducted by a researcher using an interview-based approach. Therefore, participants may have been compelled to think and perform harder during the study. On the other hand, at times we failed to convince the engineers to apply certain guidelines into a given IoT application scenario. In real world situations, these PbD guidelines would need to be used by engineers without supervision (or assistance). By taking these factors into consideration, we designed a second study aimed at exploring the engineers' mindset towards the Pbd guidelines. More specifically, we explored what software engineers think about each guideline, their reasoning and decision making process when applying them. It is important to note that the data gathered in Study 2 addresses the same question as Study 1 (Round 3), albeit in a different context.  We used Study 2 to strengthen the findings of Study 1 as well as to reach theoretical saturation \cite{Lewis-Beck2004} and we will compare these results in Section \ref{sec:Discussion_and_Lessons_Learned}.}


\subsubsection{Recruitment}
\begin{figure*}[!h]
	\centering
	\includegraphics[scale=0.55]{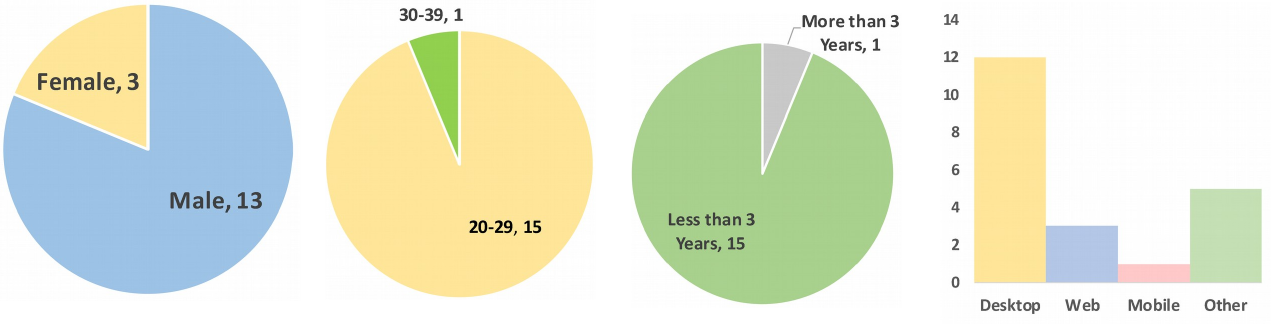}
	\caption{Demographics of Study 2 (Secondary Study)}
	\label{Figure:Demographics}	
\end{figure*}

\textcolor{olive}{In total, we recruited 17 participants, one of whom dropped out, giving us a final set of 16 participants. This survey, which was conducted at a French University with participants who were Masters students and were recruited using a convenience sampling approach. No compensation was given to the participants. The study involved completing 32 IoT use case scenarios. Based on the lessons we learned from Study 1, we did not consider the level of experience to be a relevant factor in this study.  The demographic summary for the participants is presented in Figure \ref{Figure:Demographics}.}

\subsubsection{Procedure}
\textcolor{olive}{This study was organised using two online surveys. Each participant was given three days to complete the activity. As in Study 1, we used the three case studies presented in Section \ref{sec:Motivation}. We formulated the each survey into two logical rounds (in contrast, to the three rounds in Study 1): \textit{1) without privacy guidelines}, and \textit{2) with privacy guidelines} the details of which are as follows:}

\textcolor{olive}{\textbf{Round 1:} A use case scenario is presented to each participant. Then, we asked the question \textit{``What kind of privacy protecting measures might you incorporate into the IoT application design?''}. We also recommended the participants to sketch a data flow diagram saying \textit{``Even though it is not required, it might be useful for you to sketch a data flow diagram to understand how you might want design the IoT application''}.}

\textcolor{olive}{\textbf{Round 2:} In this round, we presented different PbD guidelines, one by one, and asked the participants to answer appropriately. ( \textit{``Please read the above privacy guideline. Do you think this guideline can be applied to the IoT application design? If 'Yes'; please briefly explain how you might apply this guideline. If 'No': Please explain why this guideline cannot be applied'').}}

\section{Findings, Discussion and Lessons Learned}
\label{sec:Discussion_and_Lessons_Learned}

\textcolor{Mahogany}{In this work, we followed the multimethod-multistrand method \cite{Tashakkori2010}. More specifically, we used two data collection methods (i.e., interviews and online activities) and collected multiple types of data (i.e., IoT application designs [\textit{drawings}]), participants views [\textit{audio}], participants ability to identify privacy preserving measures [\textit{numeric}]).} In this section, we first analyse and discuss the results quantitatively. Our aim is to address the three questions presented earlier in Section \ref{sec:studey1} with the help of data collected through Study 1. Later, we discuss the results of both Study 1 and 2 qualitatively in order to understand software engineers' approach towards designing privacy-aware IoT applications. 

\subsection{Quantitative Analysis (Exploring Effectiveness)}

\begin{figure}[!h]
	\centering
	\includegraphics[scale=0.67]{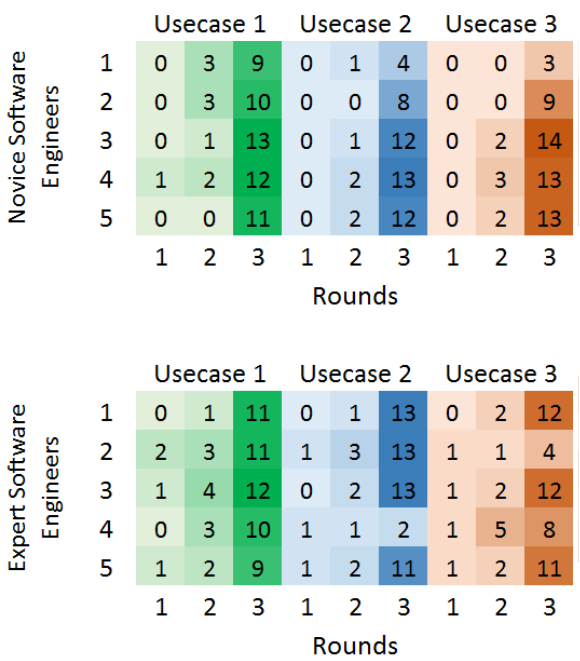}
	\caption{The three use-cases are marked using three separate colours. The x-axis denotes how many privacy protecting measures have been identified in each round (the darkness of the sharing is proportional to the number of privacy requirements identified). The  y-axis denotes the participant ID.}
	\label{Figure:ResultsHeatMap}	
\end{figure}

As shown in Table \ref{Tbl:Privacy_Requirement_Extraction}, in Study 1 we expected each participant to identify a maximum of 41 privacy protecting measures (Use-case 1: 12 measures, Use-case 2: 14 measures, Use-case 3: 14 measures). The participants may identify these privacy measures either using their experience, common sense, or using the PbD guidelines. In total, we collected 410 data points (41 measures x 10 participants). We present an overview of the data gathered using two heat-maps in Figure \ref{Figure:ResultsHeatMap} where the results for novice and expert software engineers are presented separately. 

\begin{figure}[!t]
	\centering
	\includegraphics[scale=0.50]{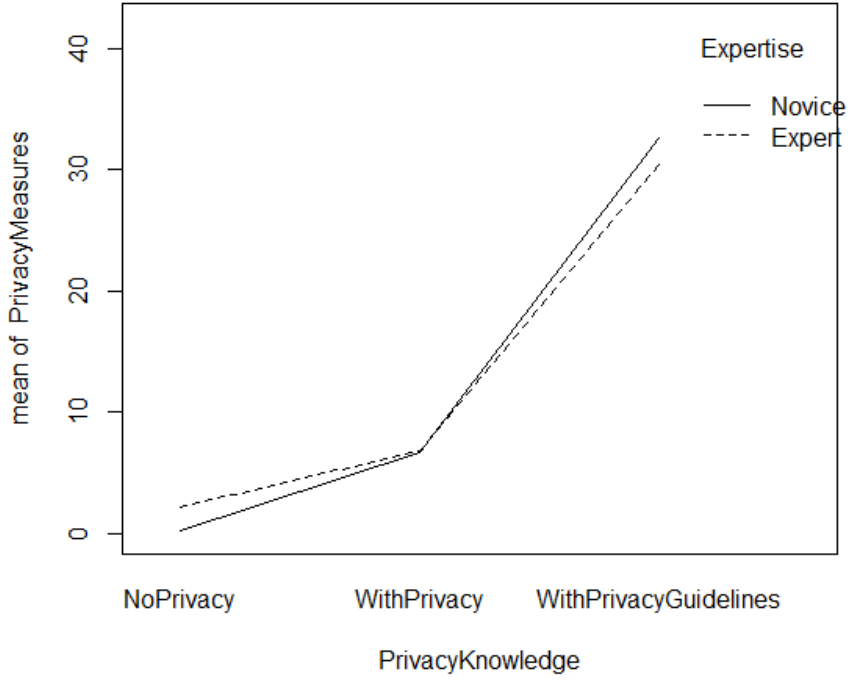}
	\caption{{Number of privacy measures} identified in each round}
	\label{Figure:Differences}	
\end{figure}

\begin{figure*}[!b]
	\centering
	\includegraphics[scale=0.45]{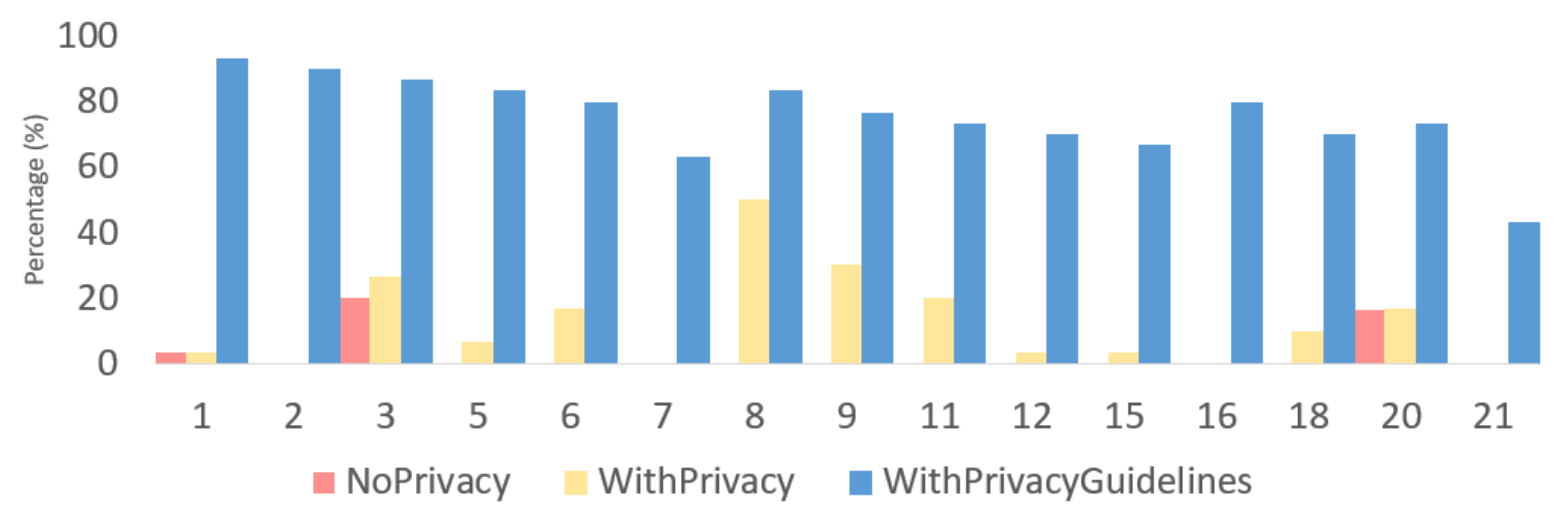}
	\caption{Study 1 - Privacy guidelines identified in each round: the x-axis denotes privacy guidelines by number and each colour represents the three rounds. The y-axis denotes the frequency with which participants identified a given privacy guideline. Legend for both Figure \ref{Figure:PbDGuidelineBasedAnalysis} and Figure \ref{Figure:PbDGuidelineBasedAnalysis2}: {\footnotesize 1-Minimise data acquisition, 2-Minimise number of data sources, 3-Minimise raw data intake, 5-Minimise data storage, 6-Minimise data retention period, 7-Hidden data routing, 8-Data anonymisation, 9-Encrypted data communication, 11-Encrypted data storage, 12-Reduce data granularity, 15-Distributed data processing, 16-Distributed data storage, 18-Geography based aggregation, 20-Time-Period based aggregation, 21-Category based aggregation.}}
	\label{Figure:PbDGuidelineBasedAnalysis}
	\vspace{-8pt}	
\end{figure*}

\begin{figure}[!t]
	\centering
	\includegraphics[scale=0.73]{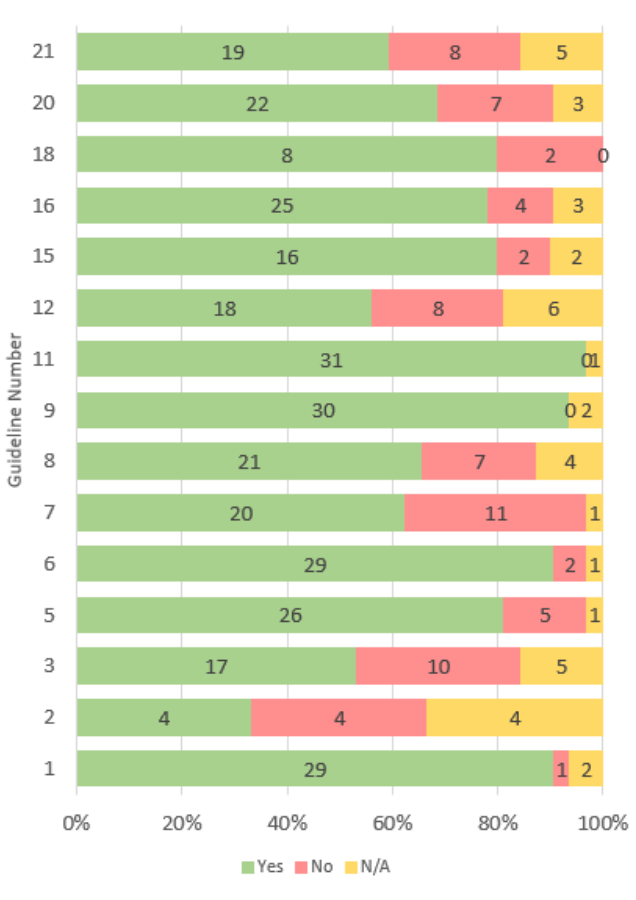}
	\caption{Study 2 - Participants' view on whether a given guideline can be applied or not to the given IoT use case scenario. Legend: Yes = participant agrees that a given guideline can be applied; No = participant refuses to apply a given guideline; N/A = participant did not clearly specify whether the guideline is applicable or not. }
	\label{Figure:PbDGuidelineBasedAnalysis2}	
	\vspace{-8pt}
\end{figure}

The heat-maps clearly show that both novice and expert software engineers were able to identify a greater number of privacy protecting measures by using the PbD guidelines than they would do otherwise. In Figure \ref{Figure:Differences}, we illustrate how the mean of the \textit{`number of privacy measures'} identified changes at different privacy knowledge levels, for novice and experts software engineers. The average number of privacy measures identified, in Round 1, by novices is \textcolor{black}{0.2} and experts is \textcolor{black}{2.2}. Similarly, the average number of privacy measures identified, in Round 2, by novices is \textcolor{black}{6.6} and experts is \textcolor{black}{6.8}. Further, the average number of privacy measures identified, in Round 3, by novices is \textcolor{black}{32.6} and experts is \textcolor{black}{30.4}.

Next, we ran statistical tests (i.e., ANOVA\footnote{statistics.laerd.com/statistical-guides/one-way-anova-statistical-guide.php}) and found out  that there is a significant difference between the number of privacy measures identified with and without the PbD guidelines (within=PrivacyKnowledge \textcolor{black}{(ANOVA p = 2.099781e-09; p $<$ 0.05)}). Further, our results show that the expertise of the software engineers (novice vs. expert) has no significant effect on the identification of privacy protecting measures (between=Expertise \textcolor{black}{(ANOVA p = 6.897806e-01; p $<$ 0.05)}) .

Figure \ref{Figure:PbDGuidelineBasedAnalysis} illustrates which privacy guidelines have been identified in each round by the participants. It is important to note that PbD guideline 2 and 18 were only relevant in one of the use case scenarios which explains its unusually low identification rate in Figure \ref{Figure:PbDGuidelineBasedAnalysis2}. To avoid any confusion, we have presented the x-axis of the Figure \ref{Figure:PbDGuidelineBasedAnalysis2} as a percentage. Comparatively, more participants have identified PbD Guideline 3 (Minimise raw data intake) and 20 (Time period based aggregation) in Round 1. However, our discussions revealed that participants integrated these features into their designs to meet functional requirements of the scenarios rather than due to a consideration of privacy. In Round 2, after we explicitly asked them to improve the privacy awareness of their IoT application designs, participants primarily identified Guideline 8 (data anonymisation), Guideline 9 (encrypted data communication), and Guideline 11 (encrypted data storage). In Round 3, there was no noticeable difference in the guidelines identified by the participants.

\textcolor{olive}{Results from both Study 1 (Figure \ref{Figure:PbDGuidelineBasedAnalysis}) and Study 2 (Figure \ref{Figure:PbDGuidelineBasedAnalysis2}) are comparable, showing that participants mostly understand and agree with the usage of encryption (communication and storage) and data minimisation very well. However, we observe a higher refusal / disagreement rate in Study 2. We discuss this phenomenon further in Section \ref{sec:Qualitative}.}


In total, we expected participants to identify a maximum of 410 privacy preserving measures that they could take in order to improve the privacy awareness of the three given IoT application scenarios. They identified  308 privacy preserving measures  with the help of the PbD guidelines, giving a success rate of 75.12\%. As shown in Figure \ref{Figure:Differences}, this result is significantly better than \textit{`without PbD guidelines'}. Based on our discussions with the participants, we identified two main reasons why they sometimes failed to apply a given guideline to their designs: 1) the IoT application designs eliminates the need to apply certain privacy preserving measures; and 2) the lack of time. The former reason arises because PbD guidelines can only be applied in certain application design contexts. Some participants designed their IoT applications such that certain PbD guidelines were not relevant. We discuss one such example in the next section.

\subsection{Qualitative Analysis and Lessons Learned}
\label{sec:Qualitative}
\textcolor{Mahogany}{We followed Miles' framework \cite{Miles2013} to conduct the qualitative analysis. Further, for data reduction phase, we use Richards' three tier coding technique (i.e., descriptive coding, topic coding, and analytic coding) \cite{Richards}. The thematic areas we found by analysing the data across both studies are as follows:}

\begin{enumerate}
	\item Challenges of the methodology and opportunities.
	\item Challenges towards adoption in the real-world.
	\item Software engineers need to develop a \textit{Privacy Mindset}.
	\item Privacy guidelines provide cues for software engineers to follow and explore beyond their expertise.
	\item Knowledge limitations and gaps could lead to weaker privacy designs.
	\item Convincing software engineers to apply PbD guidelines could be difficult.
	\item From Guidelines to Patterns: Different types of advice could be useful for software engineers to solve different privacy problems.
	\item Guidelines should be better explained.
	\item Guidelines are important and provide interesting ideas towards designing privacy aware applications.
	\item Post hoc rationalisation: Software engineers felt guilty for not pro-actively taking measures to protect user privacy.
	\item Acquisition of user consent should not be used to counter poor privacy design choices.
	\item Stimulating and supporting consistency in privacy-aware designs.
	\item Software engineers' IoT applications designs are influenced by their own expertise.
	\item Privacy should not be treated like a secondary objective when designing IoT applications.
	\item Some privacy issues can be eliminated by using alternative technologies.
	\item Software engineers consider authentication and encryption as the only ways to protect privacy.
	\item Over thinking and applications could lead to unnecessary complexities.
\end{enumerate}

\subsubsection{\textit{\textbf{Challenges of the methodology and opportunities}}}

\textcolor{Mahogany}{As shown in Section \ref{sec:Motivation}, we formulated our study based on IoT use case scenarios. During the design of this study, we had to make a decision about the level of detail that we would provide in each scenario. Our aim was to provoke the participants thought process. Therefore, we decided to keep the scenario as brief as possible. However, we wanted to give them sufficient context information to start their thinking. By doing this, we expected participants to face difficulties in designing the IoT applications without our (i.e., the interviewer's) help. Therefore, we expected them to ask lot of questions about the scenario and design requirements. Further, we always informed the participants that we are happy to provide any information that is necessary to design the application and strongly encouraged them to ask questions. Further, we intentionally embedded vague and questionable statements in each scenario to encourage participants to ask questions. Some examples of sentences from the scenarios are as follows:}

\textcolor{Mahogany}{[Extract from Use Case 2] \textit{In order to develop an efficient and effective plan, Michael needs to understand movements of people and several other aspects of their activities.}}

\textcolor{Mahogany}{[Extract from Use Case 3] \textit{However, \textit{TrueLeisure} continuously monitors and assesses the service qualities and several other aspects in each  of the amusement  parks.}}

\textcolor{Mahogany}{Ambiguous phrases such as \textit{`several other aspects'} and \textit{`understand movements of people'} forced participants to ask questions such as \textit{`What will be movements of people?'} [E1] and \textit{`What would be several other aspects? That's kind of too broad.'} [E1]. Further, in Use Case 3, we asked the participant to focus on capturing \textit{`waiting time'}. However, participant E1 challenged this by saying \textit{`Just the waiting time might not be enough'}. As expected this successfully initiated a natural conversation between the interviewer and the participant. However, none of these discussions grew into privacy requirement gathering. Participants questions were primarily directed towards functional and technological requirements. This was not a complete surprise as this kind of mindset is the challenge we are trying to address. Hence, it strengthens our argument of the necessity of providing a PbD framework that could help engineers to develop a privacy mindset.}

\textcolor{Mahogany}{It is important to note that out of ten participants in Study 1 (30 designs) only one participant explicitly discussed privacy requirements during one of the Round 1 designs. For example, when formulating a design for Use Case 2, participant [E1] said \textit{`Thinking about issues as privacy, for example, I would just be interested to know how many are there and not who is there. By that I could, for example, use the signal of the mobile phone and identify how many mobile phones are there. Then I can kind of understand the movement.'}.}

\textcolor{Mahogany}{The total duration of the activity in Study 1 was about 1.5 hrs. It is clear that we asked participants to perform a substantial task during the given time slot. Even though we did not hear any direct complaints about the workload or duration, at the time we felt that participants got tired. However, we do not consider this fatigue had any impact on our final results. On the other hand, we also need to understand that in a real-world scenario engineers would get tired. Going through privacy guidelines and deciding when, whether, or how to apply them is a significant and tiring task, especially when the number of guidelines is significant. However, if we try to reduce the number of guidelines, this will increase the abstractness and ambiguity of each guideline (e.g., Ann Cavoukian  \cite{Cavoukian}). In that situation engineer may get tired by thinking and translating principles into actionable guidelines by themselves. In either case it is hard to avoid factors that could cause fatigue when applying privacy awareness in an IoT application without building PbD support into computer-aided design tools.}

\subsubsection{\textit{\textbf{Potential challenges towards adoption in the real-world}}}

\textcolor{olive}{We observe a higher refusal / disagreement rate in Study 2 (Round 3) compared to Study 1 (Round 3). We attribute this difference to a number of factors:} 
\begin{itemize}
	\item  \textcolor{olive}{Self-administered nature of the study: Study 2 participants had 3 days to complete the task and were therefore more relaxed.  The eliminated any necessity (or pressure) to agree with the guidelines and motivated them to express their views freely.}
	\item  \textcolor{olive}{Absence of supervision: Study 2 participants completed the task on their own, so the lack of oversight for the process may have led to a lack of focus and performance.}
\end{itemize}	

\textcolor{olive}{However, Study 2 is much closer to  real world situations where software engineers have to use the proposed PbD guidelines by themselves. Therefore, tooling support will be essential to assist software engineers to improve their application designs. Automated tools will help to overcome the above two factors.}

\subsubsection{\textit{\textbf{Software engineers need to develop a \textit{`Privacy Mindset'}}}}

\textcolor{Mahogany}{Software engineers are trained to think about software designs from a business view point. This is understandable as software engineering projects typically start with business requirements conducted by business analysts. For example, in Study 1, participant [N1] recognised the importance of anonymising and deleting the data with regards to Scenario 1 in Round 2. However, he was reluctant and refused to apply the same ideas to Scenario 3 saying \textit{`I mean I can see a whole bunch of scenarios where they would want to pitch different kinds of deals to these individuals. That's why I'm saying it's very unlikely that they would adopt any sort of privacy enhancement measure, to get rid of or de-anonymise that data. Yes, just realistically I don't see that happening in that use case'}. } \textcolor{Mahogany}{N1's mindset is typical of many software engineers and while our PbD framework is a first step to changing this mindset, it is not a complete solution.  Changing engineers' thinking to make privacy a first-class consideration during design will require more effort such as sophisticated tools that can alter a given design (e.g., DFD diagram).}

\textcolor{Mahogany}{We also observed that some participants think about privacy very superficially. For example, participant [E4] was not particularly interested in thinking about privacy from certain aspects such data minimisation saying that \textit{`So, that information would, in theory, it might be possible to infer from the raw data, but in practice that could be quite tricky (Laughter).'}. It is important to understand that not all engineers understand the challenges and risks relating to privacy. For example, the participant would not have said this if they knew about state of the art (e.g., accelerometer data from a smart wrist band can be used to identify ATM pin numbers\footnote{http://uk.pcmag.com/smartwatches-1/82816/news/how-smartwatch-sensors-can-reveal-your-atm-pin}).}

\textcolor{Mahogany}{Another aspect of the mindset problem is the blind assumption about some domains. For example, one of the use case scenarios we used was from the healthcare domain. Participant [E2] correctly raised the concern that \textit{`how much is that going to impact the health plan, or the rehabilitation plan for that. If they don't have access to those, to be able to link it to the medical records. Is that going to impact the patient, if the plan is kept separate from their own doctor?'}.  This was in the context of thinking about applying the data minimisation guideline. However, more often engineers tend to assume that doctors need the most accurate data with highest possible granularity to make decisions, though it is not necessarily true in all cases. For example, participant [E3] was reluctant to apply privacy preserving measures thinking that her actions would jeopardise the medical outcomes and mentioned (in Study 1 - Round 2) \textit{`This one is quite challenging, this is the medical one, because obviously we need to use that data in order for the nurses to improve the experience in some way. So I do not know'}}. \textcolor{Mahogany}{We observed similar remarks in Study 2 as well. For example, one participant refused to apply category-based aggregation saying \textit{`It's interesting, but it lacks precision in a medical context'.}}

\subsubsection{\textit{\textbf{Privacy guidelines provide cues for software engineers to follow and explore beyond their expertise}}}

\textcolor{Mahogany}{The IoT application development process requires different types of expertise to come together to work efficiently. This is a fundamental difference between traditional web, mobile, desktop, embedded software development and IoT development. Therefore, designing privacy aware applications could be challenging, particularly when the designer does not have certain types of expertise (e.g., networking \cite{6684311}, embedded design). For example, participant [E1] highlighted  his lack of expertise saying \textit{`Yes. The main problem is the cloud itself because as the data will be going through the cloud, the data will be available for attackers or someone like that. A way just so it might be or to take a look into which cloud service we are using. The protocols and this kind of stuff because this will be really, really important. Yes. It's not my speciality, this area'}. 	Privacy guidelines can effectively educate and inform intelligent, but non specialist engineers and designers. This is an important step towards developing a privacy mindset.} 

\textcolor{Mahogany}{As a side effect, engineers may also learn to identify and respect different design requirements imposed by their colleagues who are looking at a given design from different speciality points of view (e.g., networking). Further, guidelines can also force non speciality engineers to look for speciality assistance as necessary to design more privacy aware applications. Without guidance, non speciality engineers may not know where or when to seek assistance. We heard similar expressions a few times during our studies, for example \textit{`The hidden data routing, I had not actually heard of that before, I think that is quite exciting. I think, yes, that would be good to do.'} [E3]. In another instance, participant [E2] mentioned that \textit{`The distributed data processing, I had not thought about at all to be honest I do not think but yes, I think it could definitely apply to all of these in some way. I am not sure how because I do not work in networks, or do this kind of stuff but I think that it would be good'}. These expressions convince us that, guidelines play more than the guidance role, but can effectively play an educational role. }

\subsubsection{\textit{\textbf{Knowledge limitations and gaps could lead to weaker privacy designs}}}

\textcolor{Mahogany}{Previously, we discussed the challenge of engineers not having certain expertise. We also observed slightly different types of cases where the participants incorrectly believed what they knew was correct. For example, participant [N1] mentioned that \textit{`This is volunteers, it said, so I'm assuming that at the very start of this data collection, you would start off by collecting no data about the individual. Yes, so as far as you are aware, it's just somebody. So that should be fine for that.'}. However, this is not correct. Even though we might not gather personal data initially, it could be possible to track the volunteers, if the communication is not secured through encryption.}

\textcolor{Mahogany}{In another case, participant [E1] of Study 1 mentioned that \textit{`I guess it is not necessary to encrypt and anonymise the data.'}. However, this is not true. Encryption and anonymisation techniques are designed for two different purposes. The ideal approach is to do both instead of picking one. These techniques act as two lines of defence. Encryption makes the data unreadable without authorization. However, even if a malicious party manages to decrypt the data, if the data is anonymised, the attacker will have to overcome the additional barrier of de-anonymising the data in order to cause a privacy breach.} \textcolor{Mahogany}{We found similar cases in Study 2 as well. For example, one participant mentioned that \textit{`Distributed data could not be necessary if all data is strongly encrypted'}. In reality, distributed storage and encrypted storage are two independent guidelines that can be applied together.}

\textcolor{Mahogany}{Based on these above cases, it is clear that knowledge limitations of software engineers could lead to IoT application designs with weak privacy protections for user data. The most reliable approach to address this challenges is to develop automated design tools to help the software design process.}

\subsubsection{\textit{\textbf{Convincing software engineers to apply PbD guidelines could be difficult}}}

\textcolor{Mahogany}{We realised that, at times, convincing a software engineer to apply a particular privacy guideline is difficult. For example, in Study 1, participant [N5] refused to apply `categories based aggregation guidelines', even though we were successfully able to explain it to him saying \textit{`Yes, I understood, but I don't think that we need the categories based aggregation.'}. This means we need to do more to make these guidelines more useful, but also make sure we do not push engineers to over think as we discuss in section \ref{subsub:Over_Thinking}. One of the ways to address this issue could be developing privacy patterns. Patterns are more concrete and better suited to explaining their usage in a given context more effectively than guidelines.} \textcolor{olive}{We observed similar difficulties in Study 2 as well. For example, one participant has refused to apply the data minimisation guideline by saying \textit{`No, we need precise data that we can treat , to be able to understand them'}.}

\subsubsection{\textit{\textbf{From Guidelines to Patterns: Different types of advice could be useful for software engineers to solve different privacy problems}}}

\textcolor{Mahogany}{We noted that sometimes, engineers' thinking process was just wrong. For example, participant [E1] in Study 1 said that \textit{`In this case, he needs to know which one person it is. It's important because the personal is a person  then I can't anonymise or blow it. Yes. Because this one is really a personal thing, so I think the main problem is the cloud.'}. However, this is not correct. In Use Case 1, personal information can be replaced by an identifier (for example, without using the real name. However, in this particular instance, [E1] concluded that personal data has to be retained. This kind of problem can be addressed by developing patterns. As we discussed earlier, patterns are solutions for common design problems. What we discuss here is a common problem that is not some thing unique to Use Case 1. Guidelines do have limitations on how concrete or specific they can be as they are developed with the expectation of applying to a wide range of circumstances. However, patterns on the other hand are ideal for addressing this kind of problem.}

\textcolor{Mahogany}{In a slightly different case, during Study 1 participant [N4] refused to apply the `minimise raw data intake' guideline saying that \textit{`I think it was not considered in scenario three, where I said that we will be sending the video feed to the Cloud. That can actually give the information regarding a particular user at that particular place'}. Then we asked the question \textit{`Is it necessary to send the entire video?, Is it sufficient to extract and send some pieces of that?'}. Then the participant realised the applicability of this guideline and mentioned that \textit{`Yes, instead of just sending the- because I was using the video feed- in the beginning I was using the video feed to calculate the queue times in the parking. So instead of sending just a complete video, you can just send the number plate information, if it can be done at the module at the camera. So you don't need to send that, because that will violate the personal space and privacy.'}. This situation is somewhat difficult to handle by guidelines alone. Guidelines are designed to be  broader than patterns and it is difficult to provide concrete examples. We believe that this kind of challenge can also be better managed through privacy patterns.}

\textcolor{Mahogany}{Let us consider the following extract from participant [E4]. \textit{`So, Minimised data acquisition for study one. Actually, this was an interesting one, that I would say we didn't really think about at the beginning, because one of the ways in which I have failed to minimise data acquisition is continuous data collection, (Laughter) which we did have a reason for that, which was that it might be difficult for the user to have to switch- remember to switch it on and off.'}. This problem could have easily avoided by programming the mobile devices to automatically switch on and off the data collection based on the context (e.g., when doing the exercise). However, sometimes we all may run out ideas and need a little help. Privacy patterns and automated tools could come in handy to address this type of challenge.}

\subsubsection{\textit{\textbf{Guidelines should be better explained}}}
\textcolor{Mahogany}{We also had few instances where participants struggled to understand the differences between some guidelines. For example, participant [E1] asked \textit{`What's the difference between the reduced data granularity and the minimise data acquisition'}. Such difficulties can easily be addressed by providing an example. Further, we also had some disagreements with some guideline descriptions. For example, participant [N1] mentioned that \textit{`Yes, that's what I'm guessing. Can you call it distributed processing?'}. The root cause of this problem is that, most engineers think distributed processing is all about processing at different clouds or servers. However, hierarchical data processing  also comes under distributed processing (e.g., some processing happens within micro-controllers and the further processing happens in the gateways and final processing happens in the cloud).} \textcolor{olive}{We observed similar remarks in Study 2 as well. Participants have mentioned in several places that they do not understand certain guidelines or how they can be helpful. Figure \ref{Figure:PbDGuidelineBasedAnalysis2} clearly illustrates this issue. However, these types of confusions and weaknesses (of PbD guidelines) can be easily addressed by providing examples.}

\subsubsection{\textit{\textbf{Guidelines are important and provide interesting ideas towards designing privacy aware IoT applications}}}

\textcolor{Mahogany}{Over the course of the study, a number of times, participants clearly and sincerely expressed that guidelines  are useful. For example, relating to the `minimise data retention period' guideline participant [E1] mentioned that \textit{`This, I haven't thought about it and this is very important. Very important.'}. We also had a number of instances where our guidelines have successfully changed the mind of the participants. For example, [N5] admitted the  importance aggregating data saying \textit{`So now I think if we collect the GPS data of that user, we need to aggregate the data by showing the GPS. The time periods by each aggregation, yes, I think this is quite an important thing because before that I did not think at all about that, but now I think instead of storing the raw data or the real time data, we just store the data in a certain amount of time, like, an hour or per days or per week, per month'}.}

\subsubsection{\textit{\textbf{Post hoc rationalisation: software engineers felt guilty for not pro-actively taking measures to protect user privacy}}}

\textcolor{Mahogany}{We also observed post hoc rationalisation from most of the participants. After we showed the PbD guidelines, most of the participants felt the responsibility of addressing privacy issues in their IoT application designs. Their reactions when they realised some of the privacy issues with their designs suggested that they felt guilty about initially missing them. Most of them not only followed the guidelines and successfully improved their designs, but also claimed that they thought about certain privacy considerations before we showed them the guidelines, even though their designs did not show any evidence of this. This behaviour suggests that software engineers are well aware of the importance of privacy issues, though they do not make any effort to address them until an external impetus explicitly encourages them to do so. When we explicitly encouraged them to address privacy issues, most of the participants felt the need to defend themselves and claim that they thought about privacy before. This post hoc rationalisation behaviour justifies the importance of developing a \textit{Privacy Mindset} among software engineers. We observed three different types of responses: (1) revisionist answers where the participant says that they have thought about a certain guideline, but they have not mentioned it in their designs on paper and there is no evidence to suggest that they thought about it (e.g., \textit{`So, I think I did consider the minimising the data that has been recorded'[E2]}), (2) reluctant acknowledgement that they haven't thought about it (e.g., \textit{`So, seven, I had sort of considered that, but need to make it more explicit'}[E2]), (3) reluctant acknowledgement with some guilt (expressed in facial expressions and tone) (e.g., \textit{`It is tricky actually because when you are thinking about stuff you are like I am kind of understand it, but I was not really thinking that at the time. [Laughter]. So maybe actually the walking one should be N/A as well actually'} [E3]).}

\subsubsection{\textit{\textbf{Acquisition of user consent should not be used to counter poor privacy design choices}}}

\textcolor{Mahogany}{We noticed the notion of using 'consent forms' as a way to overcome or bypass privacy challenges was a common option for many of the participants in our studies. In other words, engineers may come up with sloppy or poor application designs (in terms of privacy awareness) by using consent forms as an excuse. For example, participant [E2] mentioned that \textit{`Okay. So, the first use case. Assuming that all the patients were part of the trial that the researcher is doing, and had already signed up to allowing the data to be tracked.'}. Further, she mentioned that \textit{`The second one, as I said, these were volunteers, so, under the assumption that they've been signed up and made fully aware that this is going to track their movements'}. However, such a data collection approach is not allowed under the new GDPR regulations \cite{EUROPEANCOMMISSION2012} where all the data collection and retention activities need to be justified and documented.} \textcolor{olive}{We made similar observations in Study 2 as well. One of the participants  mentioned that \textit{`once analyses are made, data should be destroyed. However, the user may want to access to his old data to know his evolution. So I think it's not possible to destroy them, unless the user asks for this'}. Ideally, there has to be a properly justified reason in order to store data. Therefore, storing data until user explicitly asks for deletion is a weak design choice, particularly in the context of GDPR.}

\subsubsection{\textit{\textbf{Stimulating and supporting consistency in privacy-aware designs}}}

\textcolor{Mahogany}{We also noticed that some participants struggled to maintain a consistent approach across the different scenarios. For example, participant [E1] suggested using secure protocols for communication with regards to Use Case 1 even before seeing our privacy guidelines. However, he did not suggest using secure protocols for Use Cases 2 \& 3. Later, he did make the suggestion after seeing our guidelines saying \textit{`Yes. This would help with one, but with two and three, I haven't thought about it. Yes. I guess they are important to the use case two and use case three. That I haven't thought it but yes. It is really good to think about it'}. This issue is quite normal in many other domain. Maintaining consistency without any assistance is difficult. As we described in Section \ref{subsec:Why_Guidelines}, in the medical field,  check-lists were developed to guide surgical procedures. This is due to the fact that, even highly skilled doctors and medical staff struggled to always maintain consistency in practice without any assistance (i.e., reference points) \cite{Haynes2009}.}

\textcolor{Mahogany}{Additionally, we noticed that guidelines can also act as a stimulus to help engineers act upon things they already know. For example, [E3] admitted that she knows about data retention very well. However, she did not apply them in the design and said that \textit{`Yes, it is kind of in line with this one here. So, I kind of had inadvertently had thought about but probably not a mega amount. I am also the kind of person who would collect all the data and then decide to do what afterwards. [Laughter]. I am the typical scatty artist like that. With the retention period, I mean I know that it is something that you obviously need to think about, but to be honest I had not really thought about it before even this. I know from my own studies that I need to do that but when I was reading this I was not thinking, ``Oh yes, I should only keep it for a little bit.'' I guess you would delete it after you sort of put it into a secondary context.'} This means that there is a gap between engineers' knowledge and actions. Guidelines can be used to bridge that gap by helping engineers address important issues that they may not want to pay attention to.  }

\textcolor{Mahogany}{Further, guidelines can be used to eliminate the challenge of having to deal with a 'cold start' (i.e., to start thinking about something without any assistance or structure). Therefore, guidelines could speed up the process. For example, in Study 1 we had one participant who could not identify any privacy measure in Round 2 by himself. Even though this is one case out of ten, it is fair to assume this is not an isolated case. Participant [N3] mentioned that \textit{`About the privacy control, I don't have that much of knowledge about the privacy control.'} and then vaguely mentioned using policies to govern the data management process. This suggests that privacy guidelines are useful in guiding this kind of engineer}.

\textcolor{Mahogany}{We rarely noticed that participants ask direct privacy related questions in Round 1 of Study 1. Where such questions did arise, they mainly related to functional requirements gathering rather than being the result of the engineer having a privacy mindset. For example, participant [E4] asked \textit{`So, could you just give me an example of kind of sensor that we might have or an example of the sort of data that you might be collecting from one of your patients?'}. However, from a PbD perspective, a better question to ask is \textit{`What would be the minimum data set that you  need collect in order to achieve the task at hand?'}. } 

\subsubsection{\textit{\textbf{Software engineers' IoT applications designs are influenced by their own expertise}}}
\textcolor{Mahogany}{Design and development of IoT applications require different types of expertise to come together into a single design. These designs are influenced by the expertise of the engineers. For example, an engineer who is familiar with wireless network communication may look at a design with a data communications perspective. In Study 1, participant [N1] implicitly thought about data minimisation from a networking point of view \textit{`Are you gathering a lot of data, meaning you will not be able to transmit it over a wireless network? Or is it sort of a very low-bitrate data that you can collect on the cloud and analyse later? I'm wondering if you need to do any data processing at all?'}. It is important to note that engineers may implicitly apply certain guidelines without thinking about privacy, instead thinking about challenges in their own expert areas as shown above. We don't view such questions and decisions to be supportive of PbD because they are not based on privacy concerns.}

\textcolor{Mahogany}{Having expertise (or confidence) could also help engineers to make more concrete design decisions. For example, participant [N2] based on his own expertise mentioned that \textit{`In this activity, we don't need very specialised data. I think two sensors are enough, the gyroscope. I have written the gyroscope and the heart-rate monitor. That actually tells us a lot.'} In this extract, our participant, implicitly focused on the data minimisation guidelines. In this context, the participant is confident that particular data types are sufficient to satisfy the requirements. This is in contrast to the view we saw in Section \ref{Privacy_Secondary_Data_Primary}, where the participant mentioned her willingness to gather data \textit{`just in case'}. More technical knowledge and expertise of the technology could lead to a change in mindset from gathering all data to gathering sufficient data.}

\textcolor{Mahogany}{One of the ways to address this challenge is to create IoT knowledge bases as IoT application development becomes a mainstream endeavour. Therefore, it would be useful to develop usable tools that can inform engineers specifically regarding \textit{`what can be achieved by different types of data'}. For example, what can be understood by analysing accelerometer data? What can be understood by temperature data or the questions would be, What are the different ways to detect human presence in a certain locations. Different IoT application designs that achieve the same overall goal may have different consequences in terms of cost, accuracy, replicability, privacy awareness and so on. We propose to develop an IoT knowledge base  where anyone can search for answers to questions similar to the examples we provide above. Such a platform should be a crowd-sourced platform where different experts get to submit their experiences and also provide facilities to critique each others solutions. Such a resource would help the IoT application development community to collectively achieve their desired functional objectives in a privacy aware manner.}

\textcolor{Mahogany}{Privacy guidelines can also be used to justify or contrast other design decisions. For example, a decision to collect less data in order to save bandwidth can be strengthened by the arguments brought in by the data minimisation privacy guideline. Such a triangulated decision will have much better chance in surviving in design reviews by multiple parties who have different expertise. These arguments may also be useful in strengthening the justification for design decisions. For example, it would be more credible to put emphasis on the secondary benefits of the data minimisation guidelines, when possible, as it could be seen as not only a  privacy protection measure but also a cost saving measure for the company in the long run. However, the challenge is to combine privacy guidelines with secondary benefits. The knowledge bases discussed above could be  useful to address this challenge.}

\subsubsection{\textit{\textbf{Privacy should not be treated like a secondary objective when designing IoT applications}}}
\label{Privacy_Secondary_Data_Primary}
\textcolor{Mahogany}{Our study showed that software engineers do not consider privacy as a first class citizen in their IoT application designs. This justifies our decision to develop a PbD framework to guide the thought process of software engineers. During our user studies, participants candidly expressed their wish to collect as much  data as possible (e.g., Participants [E2] said \textit{`As a developer, you just want all of the data'}). We believe that this mindset of collecting as much data as possible needs to be changed towards a \textit{privacy mindset} where only the most essential data items are gathered and processed. We explained the privacy risks of gathering non-essential data in Section \ref{sec:Guidelines}.}.

\textcolor{Mahogany}{Another participant signalled that it is acceptable to collect data without any control saying that \textit{`If it's completely anonymised, and it's just business data about who's come and come out.'} [E2]. This mindset is also not supportive of PbD and creates additional problems such as resource wastage (e.g., for storage, data cleaning, data processing etc.). Further, anonymising is a risk mitigation approach, not a risk elimination approach. Anonymisation also could lead to privacy violations due to unlawful de-anonymisation approaches. We heard similar views with regards to data storage as well.}


\subsubsection{\textit{\textbf{Some privacy issues can be eliminated by using alternative technologies}}}
\textcolor{Mahogany}{An important aspect of IoT application design is the choice of the right sensors and technologies to collect data. We realised that these choices also have a direct impact on privacy. In relation to Scenario 2 (section \ref{Usecase2}), one of our participants [E4] used stationary sensors that do not capture any personally identifiable information to collect the necessary data (e.g., pressure sensors deployed in the ground, motion sensors, infra-red sensors, and so on). Sensor technologies have their own strengths and weaknesses. Similarly, privacy risks also vary depending on the technology used. However, the decision on what technology to  use is based on the exact application, associated cost, and the privacy risks that the stakeholders are willing to take. For example, deploying pressure sensors on different paths of a given park would eliminate the necessity of hiring volunteers with wearable sensor kits and the associated privacy risks. However, deploying such sensing technology in the real world could be much more challenging, in terms of cost, time, and effort, than distributing number of sensor kits among volunteers. On the other hand, stationary sensors would eliminate the hassle of recruiting volunteers, managing them, and their sensor kits. The lesson is that privacy risks can also be reduced by selecting certain types of sensing technologies providing they are feasible to be used for the particular IoT application being developed.}

\subsubsection{\textit{\textbf{Software engineers consider authentication and encryption as the only ways to protect privacy}}}
\begin{figure}[!h]
	\centering
	\includegraphics[scale=0.93]{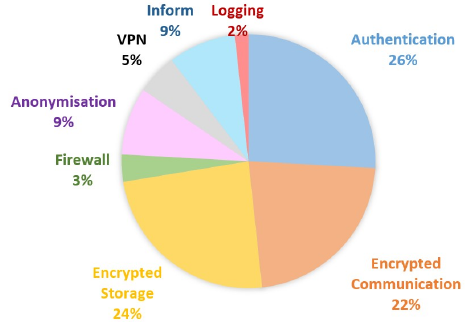}
	\caption{Common privacy protecting measures suggested by participants in Study 2 (Round 1 - Before seeing the PbD Guidelines)}
	\label{Figure:CommonPrivacy}	
\end{figure}

\textcolor{Mahogany}{It is also important to note that three participants identified authentication as a measure of protecting user privacy. However, in our PbD framework, we considered authentication \cite{6767153} as a security measure rather than a privacy protection measure. Further, three participants highlighted the importance of acquiring consent from data owners before collecting data. They also pointed out the importance of giving control to the data owners so they can decide on which data to share. Both consent acquisition (information disclosure - guidelines 22) and control (guidelines 23) appeared in our PbD framework even though we did not use them in the user study.} \textcolor{olive}{Study 2 (Round 1) also highlighted the same issue. As shown in Figure \ref{Figure:CommonPrivacy}, the most common privacy protection measures identified are authentication and encryption.}

\subsubsection{\textit{\textbf{Over thinking and applications could lead to unnecessary complexities}}}
\label{subsub:Over_Thinking}

\textcolor{Mahogany}{We noticed that sometimes, using guidelines could be tricky and engineers may apply them by over-thinking the issues. Privacy guidelines are designed to guide the thinking process rather than mandatory steps that someone should follow blindly. Effectiveness  needs to be thought through before using them. For example, participant [N5] mentioned that \textit{`Distributed data processing, I did not think about this before reading the guidelines. For scenarios two and three, we can distribute the data for processing. We send them to different Clouds, first of all with scenario three, like for attraction, like, we send the data for each attraction to different Cloud servers'}. Even though distributed processing is applicable in the scenario, it is not really a effective approach for Use Case 2. Attempting to employ multi-cloud processing as a way to apply distributed processing in Use Case 2 could lead to unnecessary complexity and higher costs with little contribution to  privacy protection. Therefore, it is important to assess each context carefully before applying a particular guideline.}


%

\subsection{Limitations}
\label{sec:Limitations}

Although all the participants were able to understand our proposed guidelines, it was apparent that familiarity is key to applying them in a given IoT application design in a short period of time. For our study, we printed the PbD guidelines on plain A4 sheets as a list.  However, the experience of our study participants highlighted that this type of printed list is difficult to follow and can be more time consuming to use. We believe that approaches such as \textit{Privacy Ideation Cards} \cite{Luger2015} and KnowCards\footnote{know-cards.myshopify.com} would be more effective by allowing users to quickly familiarise themselves with the guidelines.   In particular, using a colour coded, iconographic approach to represent the guidelines could help users remember them and thus leads to faster application of guidelines with less frustration.

An additional limitation of this work is that we did not consider the adaptive nature of privacy.  While some decisions about implementing privacy preserving measures can be taken at design-time, IoT applications are by nature unpredictable. As a result, the ability to adapt is an important feature in IoT applications. Ideally, IoT applications should be able to compose built-in privacy preserving techniques into a run-time configuration, that maximises the privacy protection level while maintaining the overall utility of the application.

\textcolor{red}{We would like to acknowledge that our design exercise is somewhat simplified compared to an real-world industrial design. For example, most our participants omitted latest distributed system design strategies such as Software Defined Networks (SDN) and Network Functions virtualization (NFV). We would attribute this  to lack of specific knowledge of our participants. However, we do not believe this issue impact the results we derive as our objective was to measure the their PbD skills, not IoT infrastructure design skills.}

\section{Related Work}
\label{sec:Related_Work}

\textcolor{purple}{Our objective is to explore ways in which we can help software engineers to efficiently and effectively design privacy aware IoT applications. Towards this, in this paper, we proposed a Privacy-by-Design framework based on a set of guidelines and an associated method for applying them. There are a number of existing frameworks that have been proposed to help elicit privacy requirements and to design privacy capabilities into systems. Privacy principles, privacy strategies, privacy patterns have been developed to support software engineering processes. It is important to note that none of these approaches explicitly focus on the IoT domain or IoT application development processes. }

\textcolor{purple}{Spiekermann \cite{Spiekermann2012} has identified a number of challenges in PbD approaches. Spiekermann identified PbD as \textit{``an engineering and strategic management approach that commits to selectively and sustainably minimise information systems' privacy risks through technical and governance controls.''}.  Privacy is a vague concept without a rigid definition. Therefore, at times, it is difficult to measure the effectiveness or efficiency of privacy protection techniques. Further, distinguishing privacy from security is vital in order to develop methodologies to address privacy challenges. Spiekermann \cite{Spiekermann2012} also highlights the problem of not having widely agreed methodology for systematic engineering of privacy into systems. This justifies our attempt to develop a methodology to incorporate privacy protecting measures into IoT application designs.}

\textcolor{purple}{Primarily, there are two approaches to incorporate privacy measures into a system design. The first approach is \textit{threat-focussed}, explicitly examines a given system design to identify privacy threats and address them. LINDDUN \cite{LINDDUN}, which we  discuss later in this section, can be considered to be an example of a threat-focussed approach. Privacy Impact Assessment (PIA)  \cite{Wright2012} is also an example of this approach. The second approach is \textit{threat-agnostic}, which proposes applying a series of privacy protecting measures to a given design without explicitly considering specific privacy threats. The expectation is to apply a set of blanket measures aiming to improve the  overall privacy awareness of the design, not worrying about the threats involved. Our proposed methodology is an example of a threat-agnostic approach. Some other examples are privacy principles \cite{Cavoukian}, and privacy strategies \cite{Hoepmanraey}. Both \textit{`Threat-focussed'} and \textit{`Threat-agnostic'} approaches have their own strengths and weaknesses. Due to unique characteristics of each approach, a hybrid approach may potentially create better system designs.}

\begin{description}
	
	\item[Threat-focussed] \textcolor{purple}{This approach eliminates specific threats that a system might have. Therefore, it is a mission oriented approach where it forces system designers to think deeply about specific threats. On the down side, systems may struggle to handle threats that the designers haven't though about during design time. Deep thinking processes would take more time and complexities could lead to poor threat analysis.}
	
	\item[Threat-agnostic] \textcolor{purple}{This approach is somewhat simpler and less  error  prone due to the absence of a threat analysis process. However, the same reason could lead to weak privacy design caused by not handling specific threats unique to a given system. On the other hand, this approach has more chance to handle unexpected privacy risks at run time due to lower dependence on a threat identification processes. Therefore, highly dynamic systems may benefit from this approach.}
\end{description}

\begin{table*}[t!]
	\caption{Summary of PbD Evaluation Methodologies}
	\renewcommand{\arraystretch}{1}
	\footnotesize
	\begin{tabular}{p{1.8cm}  p{9.5cm}}
		Area & Descriptions of Evaluation the approach  \\ \hline \\ 
		Garde-Perik \cite{VandeGarde-Perik2006}
		&

		\textcolor{purple}{This work explores the relative importance of complying with privacy related guidelines in the context of a Health Monitoring System. A total of 50 participants were given a text scenario describing a health care system. This system does not adhere to any of the OECD guidelines. Participants were then provided with potential fixes' to the system, each of which would make it comply with one specific OECD guideline. The guidelines were presented in pairs where participants needed to pick which guideline was most important.}
		\vspace{.1cm}	
		\\ 
		
		Iachello et al. \cite{Iachello2005} &

		\textcolor{purple}{This work had developed a mobile application to conduct user studies in order to extract privacy guidelines. Those guidelines are then used to develop a second mobile application to evaluate and critique the proposed guidelines. Specific guidelines are presented later in this section.}
		
		\vspace{.1cm}		
		\\		
		
		Bellotti and Sellen \cite{Bellotti1993} & 
		
		\textcolor{purple}{This work has proposed a design framework for privacy in ubiquitous computing environments. They have  proposed eleven criteria to evaluate a given design as presented later in this section. They take each criteria and evaluate it against their sample design.}

		\vspace{.1cm}		
		\\	
		LINDDUN \cite{LINDDUN} & \textcolor{purple}{LINDDUN is a threat modelling technique that supports the elicitation of privacy threats during the early stages of the software development life-cycle. Three groups have been involved in the evaluation process (total of 8 individuals) where they were asked to create a DFD diagram for a given high level scenario description (two groups focused on a e-health system and one group focused on a smart grid system) and use it to elicit the privacy threats using the LINDDUN framework. Group discussions were used to gather the participants' experience. They analysed both the results the participants documented in their reports (discovered threats), as  well as the opinions of the participants with regard to their hands-on experience.}
		
		\begin{itemize}
			\item \textcolor{purple}{\textit{Correctness:} On average, how many threats uncovered by the participants are correct (true positives vs false positives)?}
			\item \textcolor{purple}{\textit{Completeness:} How many threats are undetected by the participants (false negatives)?}
			\item \textcolor{purple}{\textit{Productivity:} How many valid threats are identified by the participants in a given time frame?}
			\item \textcolor{purple}{\textit{Ease of use:} Did the participants perceive the methodology as easy to learn and apply?}
		\end{itemize}
		
		\textcolor{purple}{In order to explore any flaws in the LINDDUN method, the researchers asked a panel of three privacy experts to perform an independent threat analysis of a smart grid system using their own expertise. They have measured the reliability by comparing the expert designs with those of their study participants.}
		
		\begin{itemize}
			\item \textcolor{purple}{\textit{Reliability:} Is LINDDUN missing any important threats?}
		\end{itemize}

		\\
		\vspace{-0.3cm}			
		
		Rubinstein and Good \cite{Rubinstein2013} &  
		
		\textcolor{purple}{Based on a review of the technical literature, this work  has derived a small number of relevant principles and illustrates them by reference to ten recent privacy incidents involving Google and Facebook.}

		\\
		\hline
	\end{tabular}
	
	\label{Tbl:MethodologySummaryTable}
\end{table*}


\textbf{Principles, Strategies, and Guidelines:} The original PbD is a framework proposed by Ann Cavoukian  \cite{Cavoukian}, the former Information and Privacy Commissioner of Ontario, Canada. This framework identifies seven core principles by which privacy sensitive applications should be developed. These are: (1) proactive not reactive; preventative not remedial, (2) privacy as the default setting, (3) privacy embedded into design, (4) full functionality positive-sum, not zero-sum, (5) end-to-end security-full life-cycle protection, (6) visibility and transparency- keep it  open, and (7) respect for user privacy, keep it user-centric. \textcolor{purple}{Cavoukian and Jonas \cite{Cavoukian2012} have extended these principles by proposing seven more specific guidelines to build PbD systems to manage big data, namely, (1) full attribution, (2) data tethering, (3) analytics on anonymized data, (4) tamper-resistant audit logs, (5) false negative favouring methods, (6) self-correcting false positives and (7) information transfer accounting.} The ISO 29100 Privacy framework \cite{ISO2015} has proposed eleven design principles, namely, (1) consent and choice, (2) purpose legitimacy and specification, (3) collection limitation, (4) data minimisation, (5) use, retention and disclosure limitation, (6) accuracy and quality, (7) openness, transparency and notice, (8) individual participation and access, (9) accountability, (10) information security, and (11) privacy compliance. \textcolor{purple}{Wright and Raab \cite{Wright2014} has proposed to extend these ISO guidelines by adding 9 more guidelines, namely, (12) right to dignity, i.e., freedom from infringements upon the person or her reputation, (13) right to be let alone (privacy of the home, etc.), (14) right to anonymity,including the right to express one's views anonymously, (15) right to autonomy, to freedom of thought and action, without being surveilled, (16) right to individuality and uniqueness of identity, (17) right to assemble or associate with others without being surveilled, (18) right to confidentiality and secrecy of communications, (19) right to travel (in physical or cyber space) without being tracked, and (20) people should not have to pay in order to exercise their rights of privacy (subject to any justifiable exceptions), nor be denied goods or services or offered them on a less preferential basis.}

\textcolor{purple}{The Fair Information Practice Principles (FIPPs) \cite{Cate2006} proposed by the United States Federal Trade Commission is also formulated as set of guidelines, namely, (1) notice / awareness, (2) choice / consent, (3) access / participation, (4) integrity / security, and (5) enforcement / redress. Organisation for Economic Cooperation and Development (OECD) \cite{Wright2011,Oleary1995} has also proposed similar privacy guidelines, namely, (1) notice, (2) purpose, (3) consent, (4) security, (5) disclosure, (6) access, and (7) accountability. Historically, OECD guidelines are considered as a successful milestone \cite{Wright2011} where it laid the foundation for both subsequent Data Protection Directive (95/46/EC) and  General Data Protection Regulation (GDPR) \cite{EUROPEANCOMMISSION2012}. Rost and Bock \cite{Rost2011} have identified six data protection goals, namely, (1) availability, (2) integrity, (3) confidentiality, (4) transparency, (5) unlinkability, and (6) ability to intervene. Fisk et al. \cite{Fisk2015} have proposed three privacy principles, namely, (1) least disclosure [internal disclosure, privacy balance, inquiry-specific release], (2) qualitative evaluation [legal constraints, technical limitations], and (3) forward progress.} 

\textcolor{purple}{Building on the ideas of engineering privacy by architecture vs. privacy-by-policy presented by Spiekerman and Cranor  \cite{Spiekermann2009}, Hoepman \cite{Hoepmanraey} proposes an approach that identifies eight specific privacy design strategies: minimise, hide, separate,  aggregate,  inform,  control,   enforce,  and demonstrate. This is in contrast to other approaches that we considered. In a similar vein, Singh et al. \cite{Singh2016a} have proposed 20 security consideration (somewhat similar to guidelines) for IoT, namely, (1) secure communications, (2) access controls for IoT-cloud, (3) identifying sensitive data, (4) cloud architectures: public, private, or hybrid?, (5) in-cloud data protection, (6) in-cloud data sharing, (7) encryption by \textit{`things'}, (8) data combination, (9) identifying \textit{`things'}, (10) identifying the provider, (11) increase in load, (12) logging at large scale, (13) malicious 'things'-protection of provider, (14) malicious 'things'-protection of others, (15) certification of cloud service providers, (16) trustworthiness of cloud services, (17) demonstrating compliance using audit, (18) responsibility for composite services, (19) compliance with data location regulations, and (20) impact of cloud decentralization on security.}

\textbf{Frameworks:}  LINDDUN \cite{LINDDUN} is a privacy threat analysis framework that uses data flow diagrams (DFD) to identify privacy threats. LINDDUN focuses on eliminating set of pre-identified privacy threats using a systematic review of data flow diagrams.  It consists of six specific methodological steps: (1) define the  DFD, (2) map  privacy  threats  to  DFD elements, (3) identify threat scenarios, (4) prioritize threats, (5) elicit mitigation strategies, and (6) select corresponding privacy enhancing technologies. However, both LINDDUN and Hoepman's framework are not aimed at the IoT domain. Further, they not prescriptive enough in guiding software engineers. \textcolor{purple}{Bellotti and Sellen \cite{Bellotti1993} have proposed a framework for design for privacy in ubiquitous computing environments. They argue that systems must be explicitly designed to provide feedback and control about (1) capture [when and what information collected], (2) construction [what happens to information], (3) accessibility [which people and what software have access to information], and (4) purposes [why data is being collected]. They also propose eleven criteria to evaluate a given design, namely, (1) trustworthiness, (2) appropriate timing, (3) perceptibility, (4) unobtrusiveness, (5) minimal intrusiveness, (6) fail-safety, (7) flexibility, (8) low effort, (9) meaningfulness, (10) learnability, (11) low cost.} In contrast, the STRIDE \cite{Stride} framework was developed to help software engineers consider security threats, is an example framework that has been successfully used to build secure software systems by industry. It suggests six different threat categories: (1) spoofing of user identity, (2) tampering, (3) repudiation, (4) information disclosure (privacy breach or data leak), (5) denial of service, and (6) elevation of privilege.  However, its focus is mostly on security than privacy concerns.

\textcolor{purple}{\textbf{Patterns and Anti-Patterns:} Both patterns and anti-patterns are important and relevant to our work. However, due to space limitations, we do not review the pattern literature in detail. Some important information on privacy patterns with relevant examples can be found at \cite{privacypatterns.eu,privacypatterns.org}.}

\textbf{Design Aids:} In a similar direction, Luger et al.  \cite{Luger2015} aims to understand how to make emerging European data protection regulations more accessible to the general public using a series of privacy ideation cards. They have extracted 40 design principles by examining the EU General Data Protection Regulation 2012 Com Final 11 \cite{EUROPEANCOMMISSION2012}.  These high level principles are proposed for computer systems in general but are not prescriptive enough to be adopted by software engineers to design and develop IoT  applications. \textcolor{purple}{In addition to using descriptions to explain guidelines, Zevenbergen et al. \cite{Zevenbergen2013} have produced a set of questions to explicitly guide the designers' mind towards following the guidelines. Inspired by their approach, the guidelines we adopted in this paper also used a question based format \cite{Perera2017b}.}

\textcolor{purple}{\textbf{Domain Specific:} Privacy guidelines can also be domain focused or contextual as well. Iachello et al. \cite{Iachello2005} haVE proposed privacy guidelines for social location disclosure applications and services. Their proposed guidelines are (1) don't start with automation, (2) flexible replies, (3) support denial, (4) support deception, (5) support simple evasion, (6) start with person-to-person communication, (7) status/away messages, (8) operators: avoid handling user data, (9) power relationships, (10) user characterization, (11) account for long learning curve, and (12) account for specific circumstances. Gritzalis et al. \cite{Gritzalis2005} has proposed 36 guidelines, formulated as counter measure, to address common privacy risks in healthcare domain. guidelines are extracted through a use case analysis and a risk assessment. Langheinrich \cite{Langheinrich2001} has developed six principles for guiding system design, based on a set of fair information practices common in most privacy legislation in use today: notice, choice and consent, proximity and locality, anonymity and pseudonymity, security, and access and recourse. Langheinrich discusses these generic principles in the context of ubiquitous computing in detail. It is important to note that, due to their abstract nature, privacy principles can be interpreted in different ways related to different contexts. Therefore, both privacy principles as well as different interpretations are both important. Cavoukian \cite{Cavoukian2006} has proposed several privacy guidelines to serve as privacy \textit{`best practices'} guidance for organizations when designing and operating Radio-Frequency Identification (RFID) information technologies and systems. The proposed guidelines are (1) accountability, (2) identifying purposes, (3) consent, (4) limiting collection, (5) limiting use, disclosure and retention, (6) accuracy, (7) safeguards, (8) openness, (9) individual access, and (10) challenging compliance. Zevenbergen et al. \cite{Zevenbergen2013} has proposed specific set of guidelines to measure mobile connectivity in a ethical way. The aim of their guidelines is to help network researchers navigate the challenges of preserving the privacy of data subjects, publishing and disseminating datasets, while adhering to and advancing good scientific practice.}

\textcolor{purple}{Cavoukian \cite{CACMStaff:2012} argues the important of empowering software engineers to develop and adopt privacy best practices. We believe that providing methodologies, tools, and techniques is part of the empowerment process.}

\subsection{Privacy Guidelines \& GDPR}
\textcolor{blue}{General Data Protection Regulation (GDPR) is a regulation enacted by the \textit{European Parliament and Council} which aims to regulate how personal data of EU citizens should be handled by any entities within or outside EU. GDPR primarily aims to give control back to citizens and residents over their personal data. This regulation is expected was adopted and implemented across the European Union in May 2018. Even though our PbD framework is not designed to specifically address GDPR, we would like to briefly highlight that parts of the GDPR regulation is organised as principles which are quite similar to the principles we discussed in this paper. An example principle is listed below.}

\begin{itemize}
	\item \textcolor{blue}{\textit{``adequate, relevant and limited to what is necessary in relation to the purposes for which they are processed (`data minimisation');''}}
\end{itemize}

\textcolor{blue}{Our privacy guidelines (especially the ones that target minimisation) will help to implement this principle. It would be useful to develop more concrete guidelines, patterns and tactics to address each of the principles proposed in GDPR.}

\section{Conclusions and Future Work}
\label{sec:Conclusion}

In this paper, we explored how a Privacy-by-Design (PbD) framework, formulated by combining a set of guidelines with a method for applying them, can help software engineers to design privacy-aware IoT applications. We evaluated the effectiveness of the proposed PbD framework through a use case based observational study where the participants were asked to design IoT applications to satisfy three given use cases.  \textcolor{blue}{Our objective is to show how a set of guidelines can  assist software engineers to design better privacy aware IoT applications.} According to our findings, the proposed PbD framework significantly improved the privacy awareness of the IoT applications designed by both novice and expert software engineers. Further, our results show that software engineering expertise does not matter significantly  when it comes to incorporating privacy protection features into IoT application designs. Finally, the qualitative data gathered during our studies highlighted a range of factors affecting privacy-aware IoT application design.  These included different gaps in engineers' knowledge and understanding of privacy; and limitations in our approach that affected engineers' ability to apply the PbD guidelines effectively.

In the future, we will conduct research to develop a set of privacy tactics and patterns that are less abstract than guidelines. Such tactics and patterns will help software engineers to tackle specific privacy  design challenges in IoT domain. At the moment, privacy guidelines are presented to the software engineers in plain text organised into a list. Though it is usable, in the future, we will explore how we can make these PbD guidelines more user friendly and accessible to software engineers. In particular, by using human-computer interaction techniques, we will help software engineers to efficiently and effectively browse and find relevant privacy guidelines, patterns and tactics in a given IoT application design context.

\textcolor{blue}{In the long term, we aim to change the way that the engineering community looks at privacy challenges. Privacy challenges are often considered to be time consuming and difficult to address and require significant expertise. \textcolor{red}{Specifically, we reviewed number of different privacy preserving techniques and ideas presented in the literature. It is quite a cumbersome task for a human designer to go through all possible privacy ideas and incorporate them into a given design.} Therefore, we need to develop new techniques that automatically address privacy challenges in the IoT application design process while letting engineers focus on other design challenges (e.g., interoperability, efficiency, etc). Such automated tools and techniques will not only transform application designs into privacy aware application designs, but also validate and verify them. These tools and techniques will also save significant engineering effort which would otherwise be needed to develop the required privacy expertise and apply it. Extending PbD frameworks such as ours with patterns and tactics will formulate the underlying knowledge base required for greater automation of privacy engineering for the Internet of Things.}

\section*{Acknowledgement}
We acknowledge the financial support of European Research Council Advanced Grant 291652 (ASAP) and the EPSRC PETRAS 2 (EP/S035362/1) . 
\section*{References}

\bibliography{library}

\end{document}